\documentclass[a4paper,11pt]{article}
\pdfoutput=1 

\usepackage{jinstpub} 
\usepackage{adjustbox}
\usepackage{graphicx}
\usepackage{tabularx,booktabs}
\usepackage{lineno} 
\usepackage{tablefootnote}
\newcolumntype{Y}{>{\centering\arraybackslash}X}
\usepackage{hyperref}
\usepackage{textcomp}
\usepackage[subrefformat=parens,labelformat=parens]{subfig}

\title{Design and construction of the MicroBooNE Cosmic Ray Tagger system}

\collaboration{MicroBooNE collaboration}
\begin{document}

\author[i]{C.~Adams}
\author[k]{M.~Alrashed}
\author[j]{R.~An}
\author[c]{J.~Anthony}
\author[aa]{J.~Asaadi}
\author[o]{A.~Ashkenazi}
\author[a]{M.~Auger}
\author[ee]{S.~Balasubramanian}
\author[h]{B.~Baller}
\author[p]{C.~Barnes}
% ...

\affiliation[a]{Universit{\"a}t Bern, Bern CH-3012, Switzerland}
% ...
\affiliation[c]{University of Cambridge, Cambridge CB3 0HE, United Kingdom}
% ...
\affiliation[h]{Fermi National Accelerator Laboratory (FNAL), Batavia, IL 60510, USA}
\affiliation[i]{Harvard University, Cambridge, MA 02138, USA}
\affiliation[j]{Illinois Institute of Technology (IIT), Chicago, IL 60616, USA}
\affiliation[k]{Kansas State University (KSU), Manhattan, KS, 66506, USA}
% ...
\affiliation[o]{Massachusetts Institute of Technology (MIT), Cambridge, MA, 02139, USA}
\affiliation[p]{University of Michigan, Ann Arbor, MI, 48109, USA}
% ...
\affiliation[aa]{University of Texas, Arlington, TX, 76019, USA}
% ...
\affiliation[ee]{Yale University, New Haven, CT, 06520, USA}

\abstract{
\\
The MicroBooNE detector utilizes a liquid argon time projection chamber~(LArTPC) with an 85~t active mass to study neutrino interactions along the Booster Neutrino Beam~(BNB) at Fermilab. With a deployment location near ground level, the detector records many cosmic muon tracks in each beam-related detector trigger that can be misidentified as signals of interest. To reduce these cosmogenic backgrounds, we have designed and constructed a TPC-external Cosmic Ray Tagger~(CRT). This sub-system was developed by the Laboratory for High Energy Physics (LHEP), Albert Einstein center for fundamental physics, University of Bern. The system utilizes plastic scintillation modules to provide precise time and position information for TPC-traversing particles. Successful matching of TPC tracks and CRT data will allow us to reduce cosmogenic background and better characterize the light collection system and LArTPC data using cosmic muons. In this paper we describe the design and installation of the MicroBooNE CRT system and provide an overview of a series of tests done to verify the proper operation of the system and its components during installation, commissioning, and physics data-taking.}


\arxivnumber{1234.56789} 


\emailAdd{microboone\textunderscore info@fnal.gov}

\proceeding{}

\maketitle
\flushbottom

\section{Introduction}
\label{sec:intro}
The MicroBooNE collaboration operates a 170~t~(85~t active) liquid argon time projection chamber~(LArTPC) neutrino detector~\cite{ub_detector}. The detector is located 463~m from the target in the Booster Neutrino Beam~(BNB) at Fermilab, and is the first operating detector of the Short Baseline Neutrino ~(SBN) program~\cite{sbn_proposal}. 
The MicroBooNE collaboration will measure neutrino cross sections on argon and investigate the low energy excess of electron-like events observed by MiniBooNE \cite{miniboone}. 


The MicroBooNE LArTPC~\cite{ub_detector} has dimensions of 2.56~m in the drift direction $x$, 2.32~m in the vertical direction $y$, and 10.37~m in the beam direction $z$. An electric field of 273~V/cm is applied to the liquid argon bulk to drift ionization electrons produced by charged particles from passing charged particles. The drifted electrons are detected by three anode wire planes. Two induction planes consist of wires placed at $\pm60$ degrees with respect to the vertical wires in the collection plane. Scintillation light produced by the passing charged particles is collected by 32 8-inch photomultiplier tubes~(PMT) placed behind the collection plane.

Sited at nearly the Earth's surface ($\sim$6 meters below grade with no overburden shielding), the MicroBooNE detector is exposed to a high rate of cosmic ray muons at 5~kHz~\cite{ub_detector}. Due to a relatively long readout window for collecting slowly-drifting ionization charge (2.2~ms), MicroBooNE collects signals from an average of 24 tracks and showers induced by cosmic muons in its data acquisition~(DAQ) window of 4.8~ms~per event. 
Therefore, improperly reconstructed cosmogenic muon tracks can be misidentified as $\nu_{\mu}$ interactions while muon tracks combined with $\delta$ rays, Michel electrons, and radiative energy depositions can be misidentified as $\nu_e$ interactions.





The MicroBooNE CRT system is an external sub-system designed to improve identification and rejection of cosmic muons. The system measures the crossing time and coordinates of the passing particles. The crossing time is compared to that of the internal TPC light detection system and to the beam timing. The track trajectories seen by CRT outside of the beam window are unambiguously identified as cosmic muons. The coordinates are compared to the extrapolated particle trajectories detected in the TPC. By matching CRT and LArTPC charged particle trajectories, we will identify and reject cosmic ray backgrounds and improve the significance of our physics measurements. Non-through-going cosmic muon with a fraction of its shower detected by CRT, another potential background, will be more affected by the application of cuts specific to an analysis, for example a fiducial volume cut on the LArTPC. The MicroBooNE CRT also enables new avenues for LArTPC calibrations using cosmic muons; one such study has been conducted using a previously installed small, 1m~$\times$~1m, external cosmic tagger system \cite{mucspaper}. 






The MicroBooNE CRT consists of 73 scintillating modules \cite{crt_novel} made of interleaved layers of plastic scintillating strips situated on the top, bottom, and long sides parallel to the neutrino beam. Scintillation photons induced by cosmogenic muons traversing the layers of each module are collected by silicon photomultipliers, and the signals are then digitized and read out by a customized front end board~(FEB). Using these scintillator strip signals, the geometric positions of incident or traversing muons can be reconstructed. The timing information is also recorded and used to match CRT events with MicroBooNE PMT and TPC signals. Extrapolated tracks reconstructed in the LArTPC aligning with CRT reconstructed tracks will be tagged as background cosmogenic muons and removed from events.

To realize these physics gains, we designed the MicroBooNE CRT system to meet the following requirements:

\begin{itemize}
\item{The configuration of planes and its supporting structure should enable the CRT to cover the maximum solid angle around the TPC.}
\item {The CRT tagger planes must be soundly mechanically supported in a way that allows integration with the existing LArTPC and the infrastructures at the liquid argon test facility~(LArTF). The CRT infrastructure must not compromise the low electronics noise environment of the LArTPC system.}
\item {The position resolution of the CRT system needs to match or be better than the position resolution of tracks projected from the MicroBooNE TPC outside the cryostat and the timing resolution of the MicroBooNE light collection system.}
\item{The CRT needs to achieve a high tagging efficiency.} 
\item{A precise and accurate timing correlation between CRT data and LArTPC data, which are timed on independent clock systems, must be achieved for the CRT data stream to be integrated into the MicroBooNE data stream. }
\end{itemize}

Given the planned use of external cosmic ray tagger systems in numerous future LArTPC-based experiments, such as ICARUS, SBND \cite{sbn_proposal}, and DUNE \cite{dune_proposal} for both background rejection and calibration purposes, MicroBooNE's fully-realized system serves as an example implementation that is relevant for future experiments. In this paper we describe the design, tests, construction, integration, operation, and performance of the MicroBooNE CRT, which was installed at Fermilab in July-September 2016~(Phase I) and February 2017~(Phase II), and was incorporated into physics data taking in October 2016.

In section \ref{sec:design} we describe the design of the MicroBooNE CRT. In section \ref{sec:test} we describe the tests of CRT components  during installation and commissioning. In section \ref{sec:reception} we describe the CRT installation. In section \ref{sec:performance} we describe the initial operational performance. 

\section{Design overview}
\label{sec:design}
\begin{figure}[htb!pb]
\center
\includegraphics[width=0.9\textwidth]{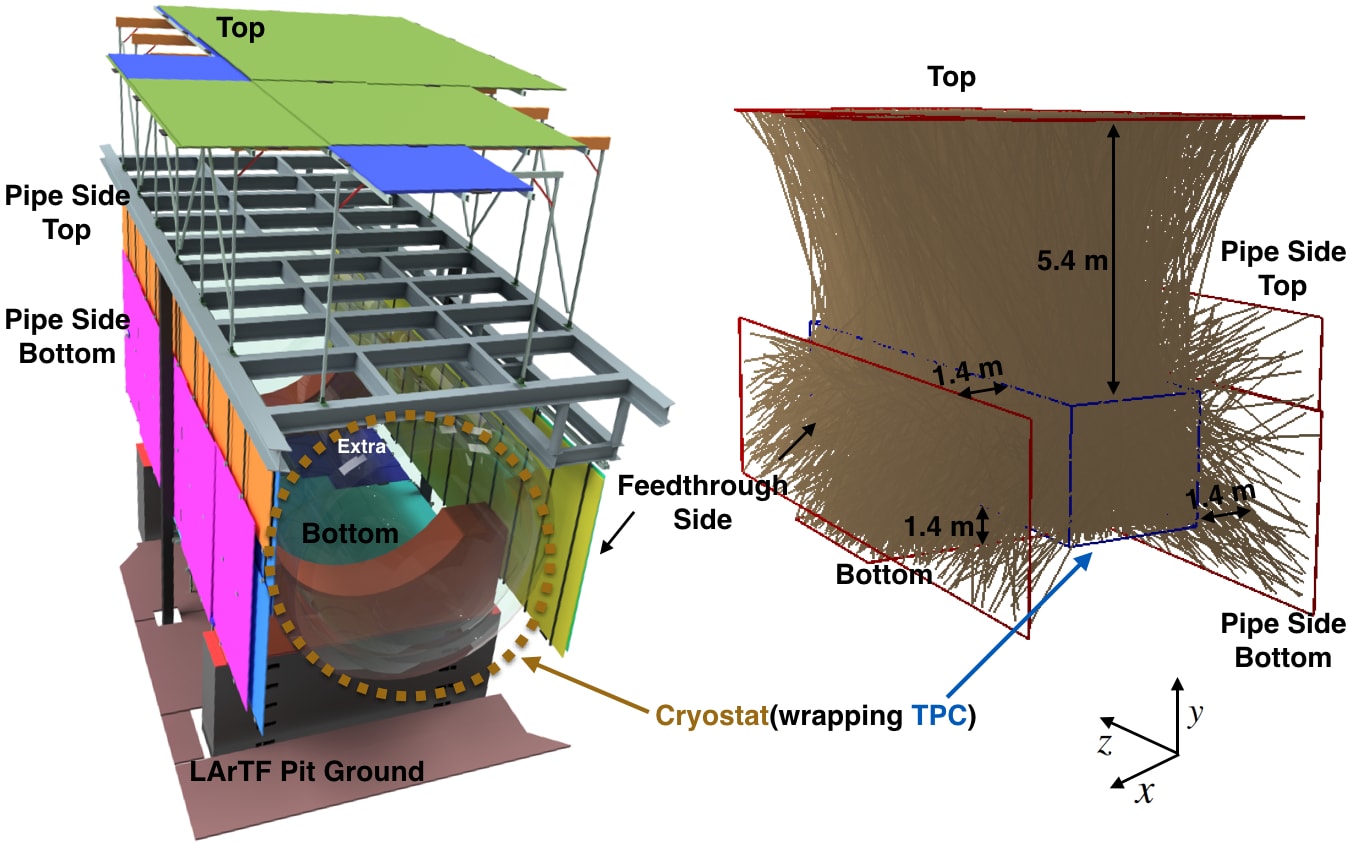}
\caption{
\label{fig:design}The design of CRT planes as part of the MicroBooNE detector. Simulation of cosmic rays crossing the CRT, the brown lines represent possible cosmic ray trajectories. There are four CRT planes: top plane, bottom plane, pipe side plane and feedthrough side plane. The beam direction is along the $z$ axis. }
\end{figure}

The MicroBooNE CRT contains four scintillating planes, each made of several scintillating modules as shown in figure~\ref{fig:design}. The scintillating module is the fundamental building unit of all CRT planes. Each module consists of a set of long scintillating strips oriented in a single common direction. The design achieves a coverage of 85\%\footnote{Cosmic rays simulated with the cosmic-ray shower library CRY, https://nuclear.llnl.gov/simulation/main.html.} of cosmic muons passing the TPC. 
In this section we describe the CRT components and demonstrate how they meet the requirements outlined in section~\ref{sec:intro}. 

\subsection{CRT planes}
\begin{figure}[htb!pb]
\center
\includegraphics[width=0.7\textwidth]{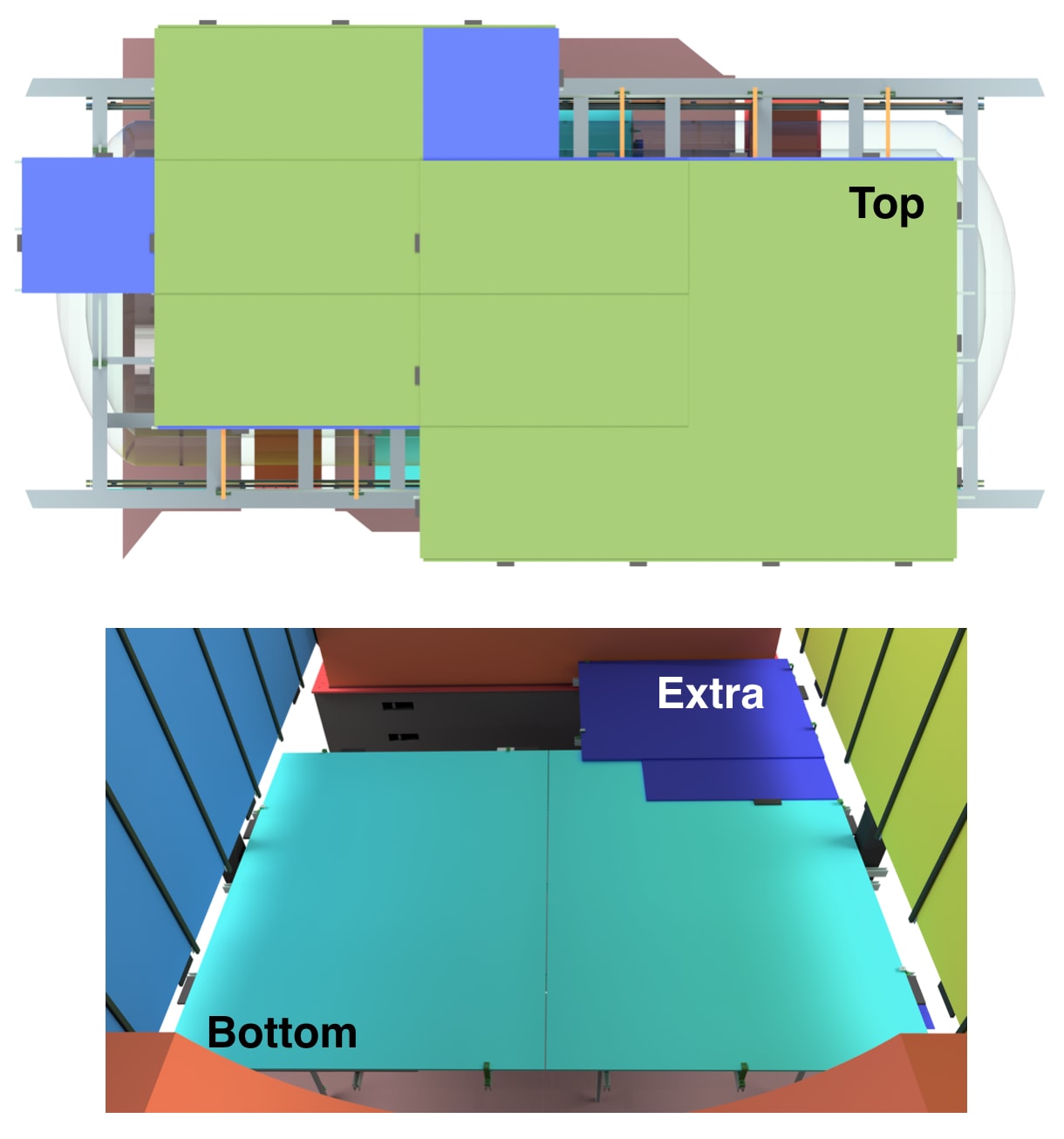}~
\caption{
\label{fig:topbottompalnes}The top and bottom CRT planes are composed of modules arranged in two layers with 24 and 9 modules respectively. In the top plane, two different sizes of modules are used (denoted in green and blue). In the bottom plane, two extra modules (denoted in blue) are used to cover the accessible corner under the cryostat. The design has reached the maximum coverage given the limited space at LArTF.}
\end{figure}

\begin{figure}[htb!pb]
\center
\includegraphics[width=0.7\textwidth]{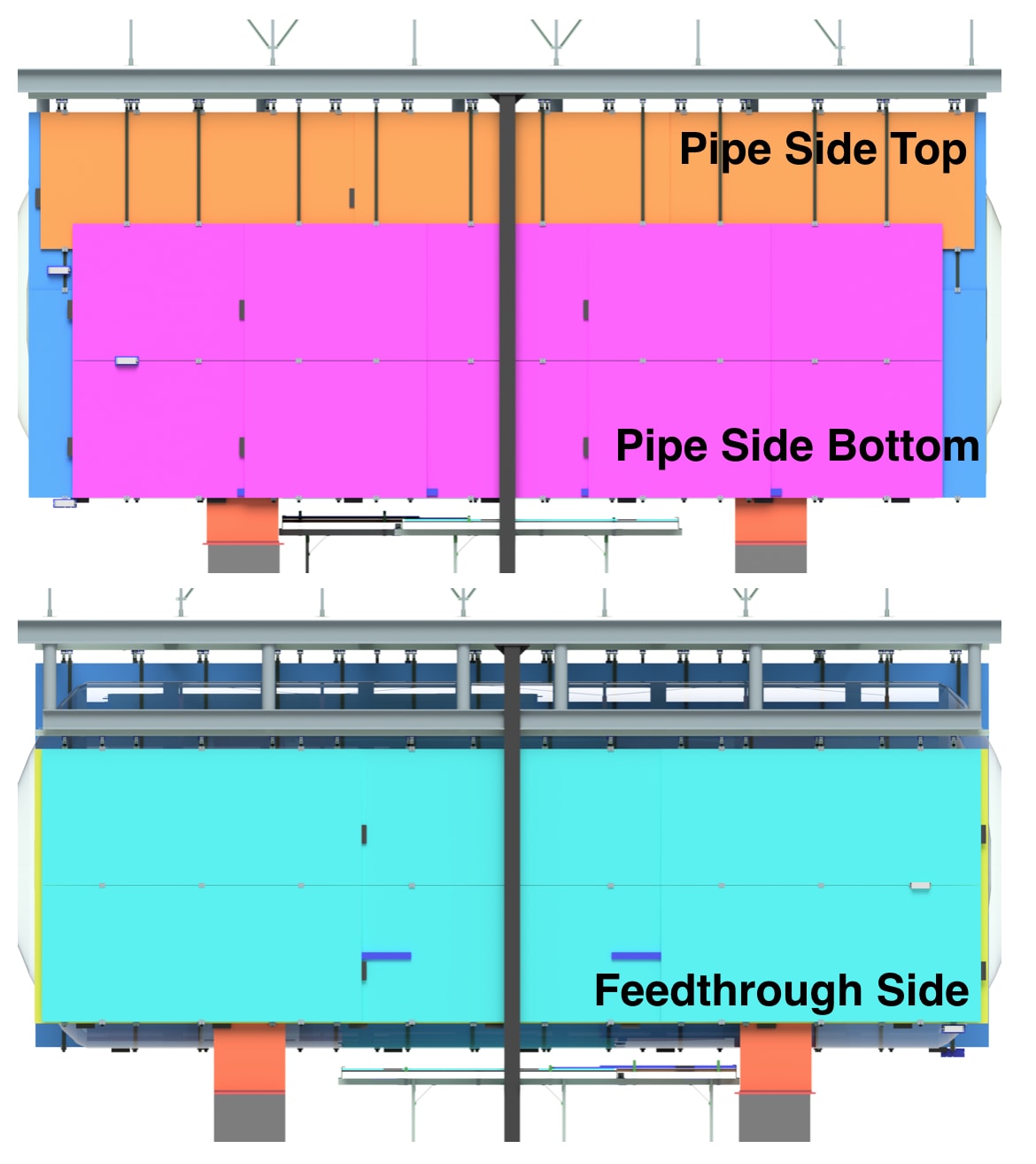}~
\caption{
\label{fig:sideplanes}The two side planes are composed of modules arranged in two layers. Layers are denoted by different colors. The horizontal layers on the pipe side plane consist of both the orange~(top) and magenta~(bottom) modules. The pipe side and feedthrough side planes consist of 30 and 13 modules, respectively. The design has reached the maximum coverage given the limited space at LArTF.}
\end{figure}

The MicroBooNE CRT consists of four planes: the top plane, the bottom plane, the feedthrough side plane, and the pipe side plane. The top plane, shown in figure~\ref{fig:topbottompalnes}, is located 5.4 m above the TPC. The bottom plane is located 1.4 m under the TPC. Figure~\ref{fig:sideplanes} shows the visible sides of the feedthrough side plane and the pipe side plane. The feedthrough side~(beam right) plane is located 1.4 m from the side where TPC feedthroughs are installed. The pipe side~(beam left) plane is located 1.4 m from the opposite TPC side. Each of the planes is composed of two layers of modules overlapping with perpendicular scintillator bar orientation. 
We describe details of each plane's construction in section~\ref{sec:reception}.

\subsection{CRT modules}

The scintillating modules were designed and fabricated by the Laboratory for High Energy Physics (LHEP) at the University of Bern. After a quality inspection and a specification test, modules were shipped to Fermilab to be installed at LArTF where the MicroBooNE LArTPC is located. More details of the construction and performance of modules can be found in Ref.~\cite{crt_novel}. Here, we summarize the most important design aspects.
 
\begin{figure}[htb!pb]
\centering
\includegraphics[width=0.8\textwidth]{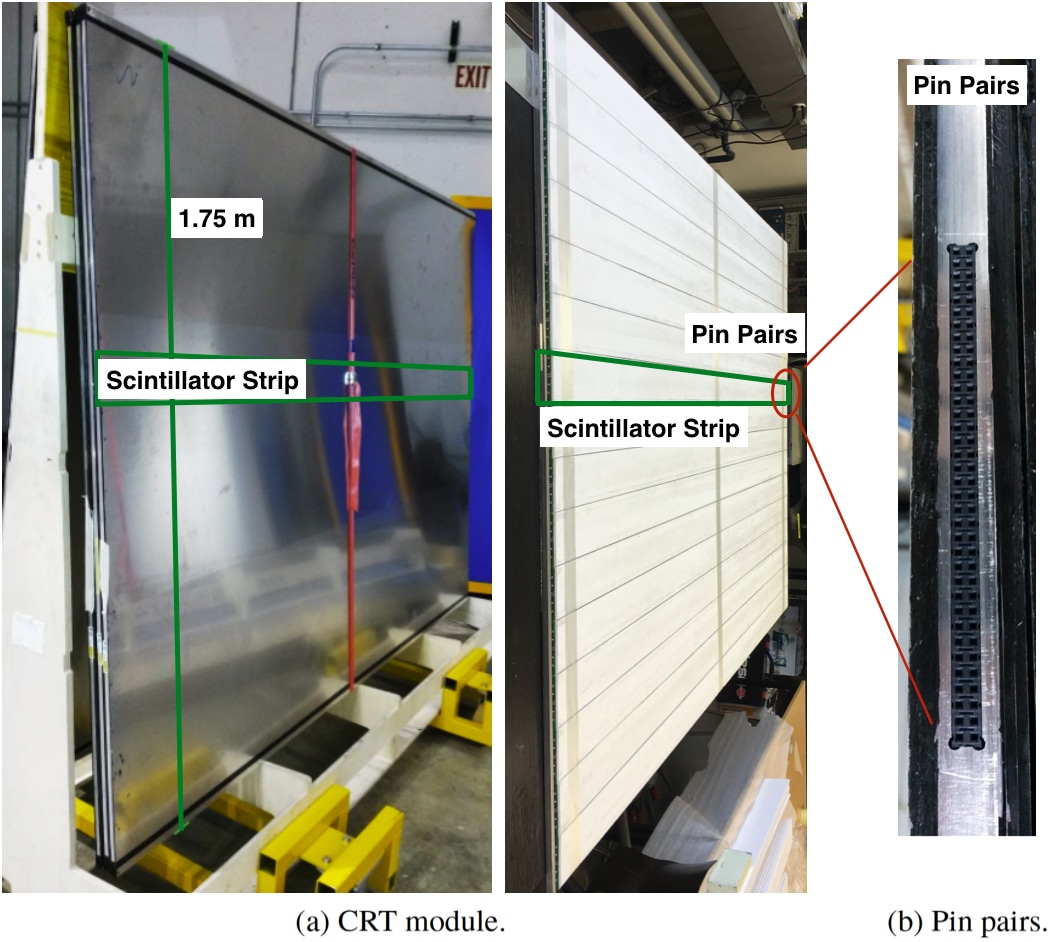}
\caption[]{(a)~CRT modules on a wooden A-frame at the DZero assembly building (DAB), Fermilab.~(b)~32 pin pairs on the readout-end for FEB for SiPM connections, plus four spare pin pairs.} 
\label{fig:panel}
\end{figure}

One scintillating module is made of sixteen 10.8~cm wide and 2~cm thick scintillating strips placed side-by-side enclosed in a 2-mm thick protective aluminum casing (figure~\ref{fig:panel}). The specifications are summarized in table~\ref{T:scintillator_strip_specs}.  To collect light signals, each scintillating strip is equipped with two WLS fibers installed in grooves along the long side. Each fiber is read out by one SiPM installed at the fiber edge. A total of 32 photosensors are deployed in each module. Modules have the same width of 1.75~m, and their lengths vary from 1.3 m to 4.1 m depending on their geometric positions. Module specifications are listed in tables~\ref{T:modules} and~\ref{T:topmoduledetails}.


\begin{table}[htb!pb]
\begin{center}
\caption{Scintillating strip specifications.}
\begin{adjustbox}{max width=\textwidth}
\begin{tabular}{| c | c | c | c | c | c | c | c | c |}
\cline{1-2}
\multicolumn{2}{|c|}{\textbf{Scintillating Strip Specifications~\cite{crt_novel}}} \\
\hline
Composition& USMS-03 polystyrene-based mixture\\ 
\hline
Emission maximum& 430~nm  \\ 
\hline
Bulk attenuation length & >7.5~cm\\
\hline
Wavelength shifting~(WLS) fiber & Kuraray Y11(200)M, 1 mm diameter\\
\hline
Cover material & Reflective aluminized mylar tape \\
\hline
Light collection system& Hamamatsu S12825-050P silicon photomultipliers (SiPMs)\\
\hline
\end{tabular}
\end{adjustbox}
\label{T:scintillator_strip_specs}
\end{center}
\end{table}

\begin{table}[htb!pb]
\caption{The assortment of modules for the bottom and two side planes of the MicroBooNE CRT, along with geometrical details and characteristics of each plane.}
\begin{center}
\begin{adjustbox}{max width=\textwidth}
\begin{tabular}{| c | c | c | c | c | c | c | c | c |}
\hline
\textbf{Section} & \multicolumn{2}{ c |}{\textbf{Feedthrough Side}} & \multicolumn{2}{ c |}{\textbf{Pipe Side Bottom}} & \multicolumn{1}{ c |}{\textbf{Pipe Side Top}} & \multicolumn{3}{ c |}{\textbf{Bottom}}\\ 
\cline{2-9}
\hline
Orientation in TPC coordinates&$z$&$y$&$z$&$y$&$z$&$z$&$y$&Extra\\ \hline
Number of strips &96 &112& 160 &224 &48 &48 &64 &32\\ \hline
Number of modules &6 &7 &10& 14& 3& 3& 4 & 2\\ \hline
Length (m) &4.04 &3.46& 2.27& 2.60& 3.96& 3.46& 2.60& 2.27\\ \hline
Mass of Front end modules (kg) &1.6 &1.9 &3.2 &3.8 &0.8& 0.3 &0.8 &1.1\\ \hline
Mass of face plates (kg) &601.8& 601.8& 676.4& 902.6& 295.2& 85.9& 257.9 &257.9\\ \hline
Mass of plastic (kg) &418.6 &418.1 &470.5& 627.9 &205.3& 59.8 &179.4 &179.4\\ \hline
Mass per module (kg) &173.9 &149.0& 98.3 &112.3& 170.6 &149.2& 149.2 &112.3\\ \hline
Mass of all modules (kg)& 1,043.4& 1,043.3 &1,179.7& 1,571.7& 511.9& 149.2 &447.6& 449.07\\ \hline
\end{tabular}
\end{adjustbox}
\end{center}
\label{T:modules}
\end{table}

\begin{table}[htb!pb]
\begin{center}
\caption{The types of modules used for the top plane of the MicroBooNE CRT, along with some of the characteristics of the plane. Two different sizes of modules are used given the access allowed by the existing infrastructure at LArTF.}
\begin{adjustbox}{max width=\textwidth}
\begin{tabular}{| c | c | c | c | c |}
\hline
\textbf{Module} & \multicolumn{2}{ c |}{\textbf{Size I}} & \multicolumn{2}{ c |}{\textbf{Size II}} \\ 
\cline{2-5}
\hline
Orientation in TPC coordinates&$z$&$x$&$z$&$x$\\ \hline
Number of strips &144& 112& 32& 96\\ \hline
Number of modules &9& 7& 2& 6\\ \hline
Length (m) &3.65& 3.65& 1.85& 1.85\\ \hline
Mass of Front end modules (kg)& 2.4& 1.9& 0.5& 1.6\\ \hline
Mass of face plates (kg)& 773.7& 601.8& 171.9& 515.8\\ \hline
Mass of plastic (kg)& 627.9& 538.2& 59.8& 149.5\\ \hline
Mass per module (kg)& 157.0& 157.0& 79.5& 79.5\\ \hline
Mass of all modules (kg)& 1,413.1& 1,099.1& 158.9& 476.7\\ \hline
\end{tabular}
\end{adjustbox}
\label{T:topmoduledetails}
\end{center}
\end{table}

A series of laboratory tests measuring the conversion gain of photosensors, the light yield of the strips, and the timing resolution and tagging efficiency of modules have been applied to all modules. The test results show a spatial resolution of 1.8~cm and a timing resolution of 1~ns for all locations on a module. 
Using an external cosmic ray telescope, the efficiency for detecting a through-going minimum-ionizing muon was measured to be >95\% at all locations on a CRT module. Details of these tests are described in Ref.~\cite{crt_novel}.




\subsection{Front end board electronics}


For the readout of SiPM arrays in the modules, a specific FEB, has been
developed by the LHEP at the University of Bern~\citep{feb}\footnote{Designed by LHEP and now commercially available from CAEN~ \cite{feb_caen}.}. 

One FEB serves all 32 SiPM channels in one module. The FEB provides bias voltage in the range of 20-90~V, individually adjustable for each SiPM channel. Amplification and shaping of each SiPM output pulse are processed and digitized individually by the FEB using a CITIROC 32-channel ASIC from Omega~\cite{citiroc} with a resolution of 12bits. 
It has a fast shaping mode with a peaking time of 15~ns and a slow shaping mode with a peaking time configurable between 12.5~ns and 87.5~ns. Signals from a fast shaper are used to provide a basic coincidence for each SiPM pair~(channels 0 and 1, channels 2 and 3, etc.) in one scintillator strip. The coincidence is formed by combining each SiPM pair routed to a field-programmable gate array~(FPGA) chip with an AND logic. The OR logic of the basic coincidences from 16 scintillating strips is then used to generate an event trigger for each module with a back-end interface. The FEB is able to generate timestamps with a resolution of 1~ns. 
For each recorded event the FEB stores two reference timestamps, GPS and beam timing, which are later used to perform the TPC and CRT event matching. The FEB is equipped with a buffer of a capacity of 1024 events and can work with triggers up to 32~kHz with a FEB latency time of 22~$\mu$s.

\subsection{CRT infrastructure}
\label{sec:rack}


We built three electronics racks for MicroBooNE CRT operation, two DAQ racks (DAQ-R1 and DAQ-R2) and one utility rack. These racks are located in the pit of LArTF underneath the cryostat. The CRT DAQ racks contain three groups of components: a rack power distribution system, a network switch, and DAQ machines. The CRT utility rack contains three groups of components: a rack power distribution system, a copper-fiber receiver and a PPS fanout for beam timing and GPS timestamps, and the DC power source for the FEBs. The contents of the racks are described below. 
\subsubsection{Rack power distribution}

For the rack power distribution system, all three racks are equipped with a rack protection system (RPS) and a slow controls box. In Ref.~\cite{ub_detector}, detailed descriptions of these components are given. Here, we summarize their necessary functions for the CRT infrastructure.

One RPS consists of a chassis connected to smoke-sensing and temperature-sensing systems interlocked with AC and DC power transmissions. The 1U RPS chassis is located near the top of the racks, along with the smoke sensor. The system is designed to meet Fermilab safety requirements and minimize the risk of fire and damage to rack components and to LArTF infrastructure. The RPS chassis produces a 12~V RPS status signal for input to a slow control chassis, also located in each rack. In the CRT DAQ racks, another 12~V interlock signal is sent to an AC Switch box located in the bottom of the racks to interlock AC power to all rack components. In the event that a rack-mounted smoke detector detects particulates above threshold, the RPS will drop the interlock signal to the connected AC switch box in the DAQ racks or the Surge-X power distribution unit~(PDU) in the utility rack, which will in turn interrupt AC current to the DC power supply. 

The slow controls chassis is part of MicroBooNE's control and monitoring system which monitors and controls the RPS, digital temperature sensors, and fan pack working states in the CRT racks. It contains a single-board computer of Glomation GESBC-9G20~\cite{glomation}, power supply, and various inputs, outputs and indicator lights. The slow controls chassis is powered from an unswitched outlet on the AC Switch Box. 

At LArTF, 208~V, 3-phase power is distributed to all CRT electronics racks. In the two DAQ racks, AC Switch Boxes are installed to distribute power to the APC Switched Rack PDU AP8932~\cite{PDU} only upon receiving an interlock signal from a smoke detection system in each rack, described above. The PDU power rating is up to 2880~W with a maximum input 30~A current. The AC switch box was designed and built at Fermilab. For the purpose of power conditioning, a commercial uninterrupted power supply (UPS) model Smart-UPS X SMX 3000 RMLV2UNC~\cite{UPS} is applied to the output of the AC Switch Box. During a fire alarm, an emergency power off~(EPO) signal will be sent out from the RPS to cut off the power on the UPS. 

With a lower power load in the utility rack, a SurgeX SX-1120-RT \cite{surgex} PDU with load rating of 12~A at 120~V is used to distribute AC power in that rack only when it is receiving the 12~V interlock signal from the RPS.





\subsubsection{Network switch}
A commercial 48-port network switch is installed in the DAQ-R2 rack to provide network connection to rack components and FEBs. Details of network connection are described in section~\ref{sec:network}.

\subsubsection{DAQ machines}
Five computer servers are installed in CRT DAQ racks as CRT server machines (3 servers mounted in the DAQ-R1 rack and 2 servers mounted in the DAQ-R2 rack). Four servers are dedicated to collecting data, with one responsible for each CRT plane. The fifth server is a spare. An event-building machine~(EVB) for CRT events is installed in the DAQ-R2 rack; fragments of data are sent from servers to the EVB over the internal network at LArTF. Full CRT events are then checked for consistency and written to local disk on the EVB before being sent offline for further processing. The power consumptions in rack DAQ-R1 and DAQ-R2 are 1350~W and 1600~W, respectively, which are about half of the PDU maximum load of 2880~W. 

\subsubsection{Copper-Fiber receiver and PPS fanout}
A GPS PPS driver chassis has been developed at the University of Bern and is mounted in the utility rack to receive GPS timestamps and neutrino beam timing signals. This module also produces clock signals that are fanned out to 20 individual outputs. Outputs are used to forward GPS and beam timing signals to each of the FEBs. The chassis DC power is produced by an AC-DC power supply, internal to the chassis, which is rated for an output of 5~VDC at 300~mA. 

The CRT racks are connected to building ground, while the non-CRT MiroBooNE racks providing GPS and beam timing signals are on the MicroBooNE platform above the cryostat on a clean detector ground, which is comparatively noise-free. To maintain separation of detector and building grounds, GPS and beam timing signals are transmitted to the CRT through a fiber connection. This connection is realized through the use of a commercial copper-to-fiber transmitter (LuxLink DT-7201)~\cite{transmitter} and fiber-to-copper receiver (LuxLink DR-7201)~\cite{transmitter} at the MicroBooNE end and CRT end, respectively. Details of GPS timestamps and beam timing connections are described in section~\ref{sec:gpsandbeamtiming}.

\subsubsection{DC power source for FEBs}
A 12-channel commercial power supply Wiener PL512~\cite{pl512} is customized and mounted in the utility rack as the source of low voltage input to the FEBs through the CRT power distribution boxes (details in section~\ref{sec:dcpower}). The Wiener PL512 power supply converts AC power to 12 independently controlled floating DC outputs. The Wiener power supply consists of a chassis and removable power unit. The Wiener chassis is wired appropriately to CRT power distribution boxes. Each Wiener output channel contains two studs for power and two compression terminals for sense lead connections. The number of FEBs connected to each channel is listed in table~\ref{T:pl512}. Each output channel is capable of providing (2-7)~V DC voltage and up to 30~A. The unit receives power from an interlock enabled Surge-X power distribution unit. To facilitate the cabling between the Wiener power supply and the power distribution boxes, the two sense and two power connections from the Wiener chassis backplane have been collected into an Andersen Powerpole $2\times2$ panel mount housing. 


\setlength{\tabcolsep}{1pt} 
\renewcommand{\arraystretch}{1}
\begin{table}[htb!pb]
\begin{center}
\caption{The different connected FEBs on Wiener PL512 low voltage power supply. Channel 10 and 11 are spare channels. }

\begin{tabularx}{\textwidth}{ |c| *{12}{Y|} }
\cline{1-13}
   \multicolumn{1}{|c|}{\textbf{Section}} 
 & \multicolumn{2}{c|}{\textbf{Bottom}}  
 & \multicolumn{3}{c|}{\textbf{Pipe Side}}
 & \multicolumn{2}{c|}{\textbf{Feedthrough}}
 & \multicolumn{3}{c|}{\textbf{Top}}
 & \multicolumn{2}{c|}{\textbf{Spare}}\\
    \multicolumn{1}{|c|}{} 
 & \multicolumn{2}{c|}{\textbf{Plane}}  
 & \multicolumn{3}{c|}{\textbf{Plane}}
 & \multicolumn{2}{c|}{\textbf{Side Plane}}
 & \multicolumn{3}{c|}{\textbf{Plane}}
 & \multicolumn{2}{c|}{}\\

\hline
Channel Number       & 0 &1 &2 &3 &4 &5 &6 &7 &8 &9 &10 &11 \\ 
\hline
Connected FEBs& 4  & 5 & 9 & 9 & 9 & 7 & 6 & 8 & 8 & 8 & 0 & 0\\ 
\hline
\end{tabularx}
\label{T:pl512}
\end{center}
\end{table}

\setlength{\tabcolsep}{6pt} 
\renewcommand{\arraystretch}{1}




\section{Performance tests}
\label{sec:test}

\begin{figure}[htb!pb]
\centering
\subfloat[Receiving and test area at DAB.]{\label{fig:testarea-a}\includegraphics[width=0.475\textwidth]{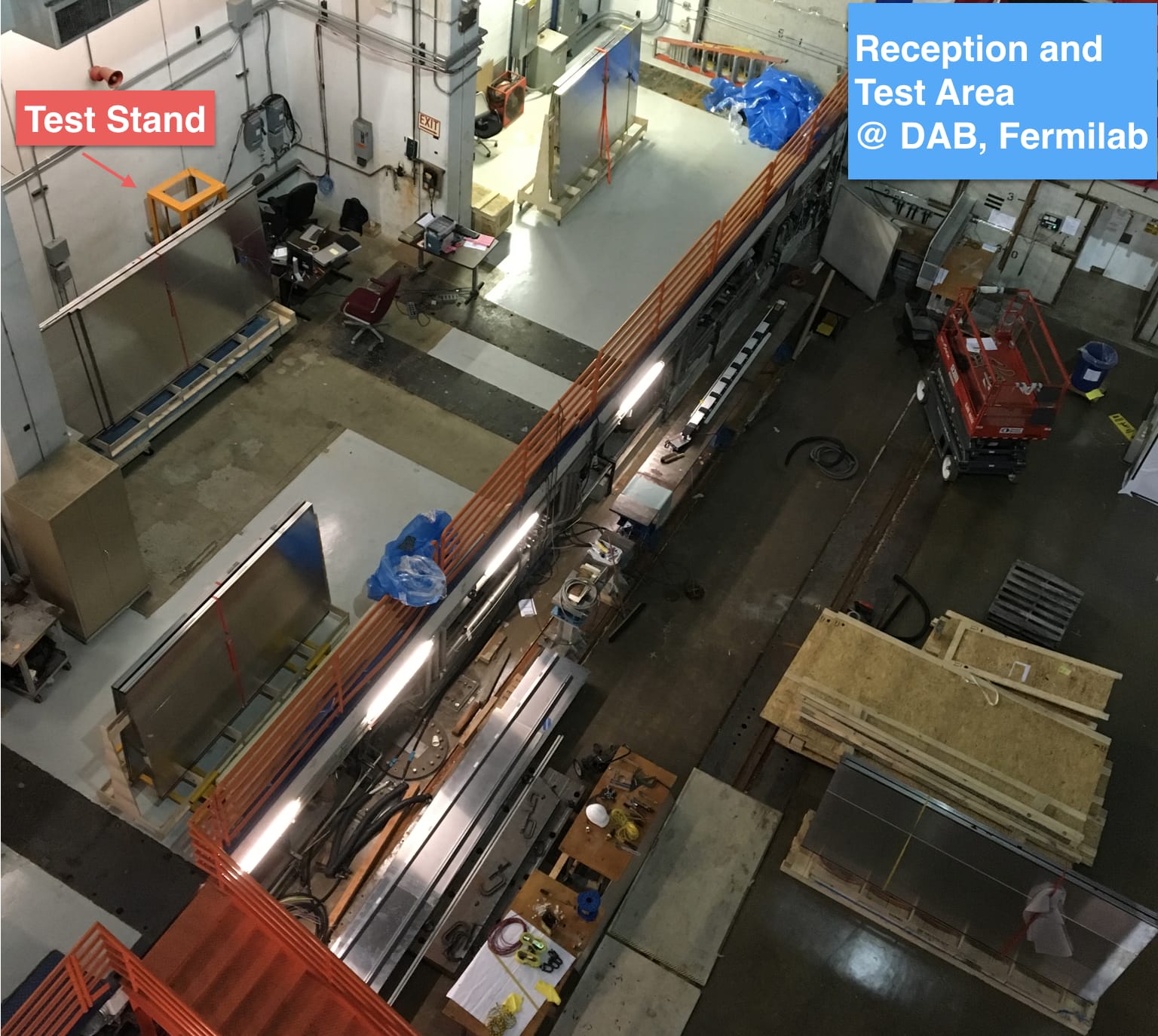}}
\hspace{\fill}
\subfloat[FEB configuration for testing.]{\label{fig:testarea-b}\includegraphics[width=0.49\textwidth]{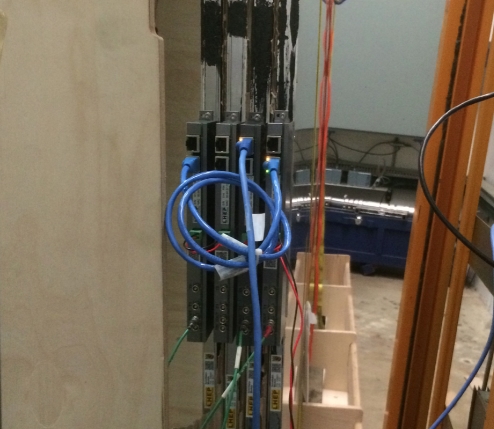}}
\caption[]{(a)~Receiving and test area for CRT modules at DAB, Fermilab. A specific test stand was built for test purposes. (b)~Example of three FEBs connected in a nested configuration. Category 5 ethernet cables were used for data transmission. LEMO cables were connected to a PPS unit in the test stand to receive the GPS timestamps. }
\label{fig:testarea}
\end{figure}

We tested the scintillating modules and FEBs before shipping them from Bern to Fermilab. The modules were shipped to Fermilab on wooden A-frames as shown in figure~\ref{fig:panel} and covered with recycled foam. After being received at Fermilab, we tested the modules and FEBs again to examine for potential damage caused to the modules during shipment. The sidewalk region in the pit of DAB, pictured in figure~\subref*{fig:testarea-a}, served as the receiving and testing area for CRT modules. Upon unpacking, we conducted a series of tests to inspect the quality and working performance of the modules in each frame. These tests verified the module quality, the module-FEB readout and triggering chain, and the general level of muon tagging efficiency. This section gives an overview of the pre- and post-installation tests of the CRT modules and FEBs.


We tested the modules on the A-frames in place, without moving them. After the tests we covered the modules with foam and packed them in frames. The frames were stored at DAB and shipped to LArTF prior to installation.



\subsection{Front end board tests}
The FEB is responsible for readout of SiPM signals, reception of timing signals, and generation of cosmic muon triggers. FEB tests were intended to examine the FEB's working states when nested (i.e. connected together via daisy-chained power, network, and timing connections). We tested each FEB for general functionality, GPS timing reading capability, and trigger rate readout as shown in figure~\subref*{fig:testarea-b}.


\subsubsection{FEB functionality}
We also tested the functionality of each FEB including the DC power and network connections. A commercial power supply HP E3610A was operated at 5.0~V/~2~A to power up to three daisy-chained FEBs. Two 19 AWG parallel wires were used to connect the power supply to the FEBs. The current draw was measured to be 1.5~A with the voltage of the SiPM arrays off and 1.6~A with the voltage on. We repeated the test for all combinations of any three modules on one side of the frame.

\subsubsection{GPS and beam timing }

\begin{figure}[htb!pb]
\centering
\includegraphics[width=0.9\textwidth]{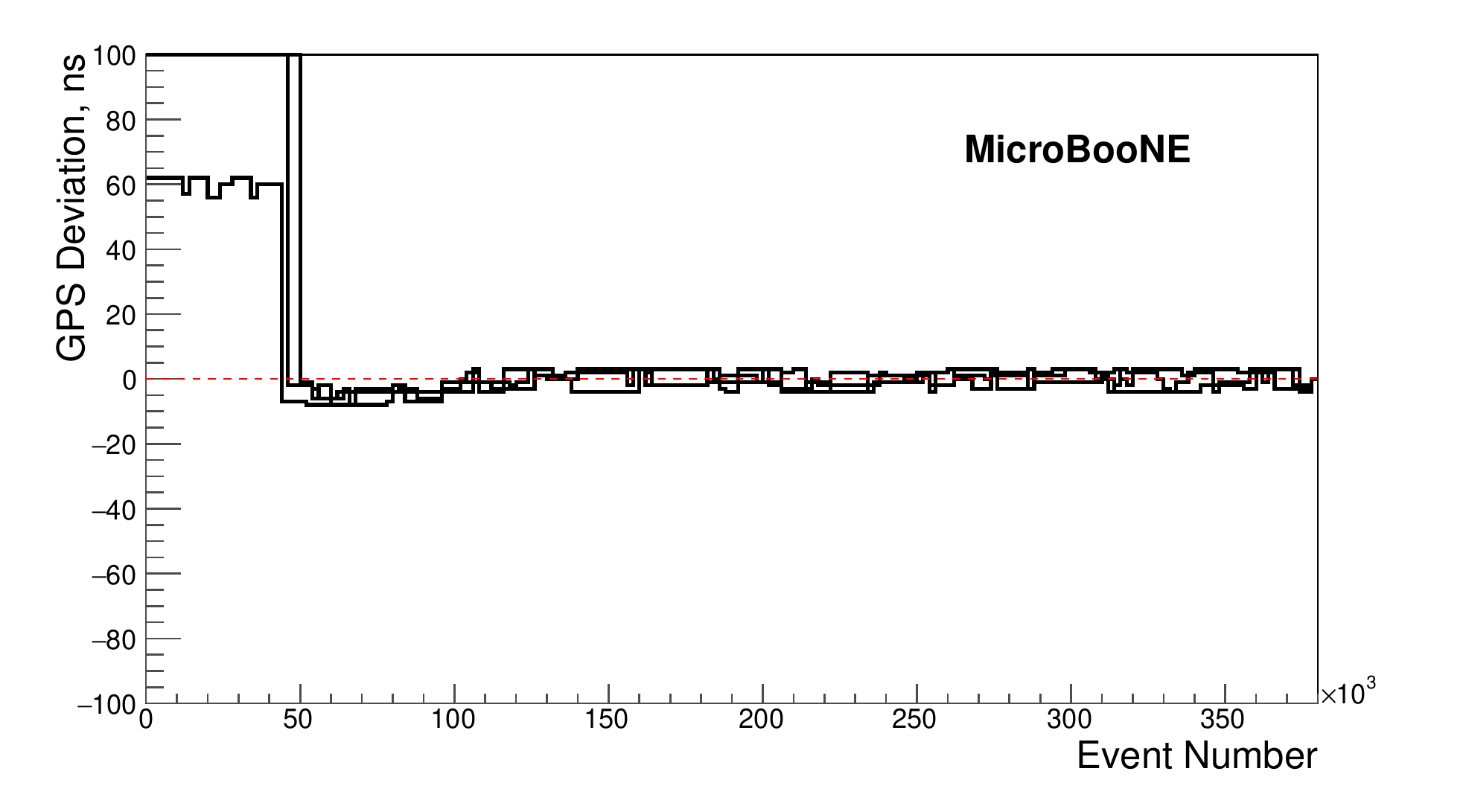}
\caption[]{Deviation of GPS timestamps of three daisy-chained FEBs. The timestamps were synchronized with a fluctuation of differences of less than 5~ns after taking $4\times 10$$^{4}$ events, which corresponds to $\sim$12 seconds.}
\label{fig:tzerotest}
\end{figure}

Each FEB has four LEMO connectors with two to receive GPS timestamps and beam timing signals, and two to generate event triggers. In this test, we examined the GPS timestamps read by different sets of daisy-chained FEBs. One commercial GPS antenna was installed on the DAB roof for GPS synchronization. The GPS timestamps were delivered in PPS format to the test stand and then fanned out to up to three connected FEBs. An example test result is shown in figure~\ref{fig:tzerotest}. GPS timestamps on connected FEBs were synchronized automatically in less than $\sim$12 seconds. 

The beam timing LEMO connector on each FEB was tested again after the CRT was fully installed at LArTF. The performance of the CRT with respect to beam timing is discussed in section~\ref{sec:crtevents}.

\subsubsection{Trigger logic}

CRT triggers are generated in the FEB. Trigger logic generation has two stages. The first stage is the scintillating strip trigger generation in each module. Each scintillating strip has two SiPM readouts. When both SiPMs detect charge above a discriminator threshold, an odd-even channel logic signal of the strip is generated and sent as output at the trigger output~(TOUT) LEMO connector. Each FEB has another trigger input~(TIN) LEMO connector to receive this TOUT trigger signal from a different FEB. In the second stage of trigger logic generation, a cross-module coincidence trigger is generated when TIN and TOUT signals are found within a 150~ns window.

\begin{figure}[htb!pb]
\centering
\subfloat[Trigger formation in production.]{\label{fig:triggers-a}\includegraphics[width=0.475\textwidth]{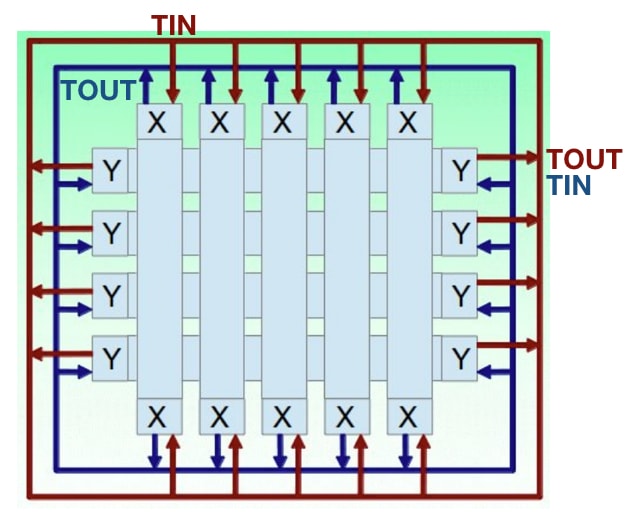}}
\hspace{\fill}
\subfloat[Trigger formation for testing.]{\label{fig:triggers-b}\includegraphics[width=0.49\textwidth]{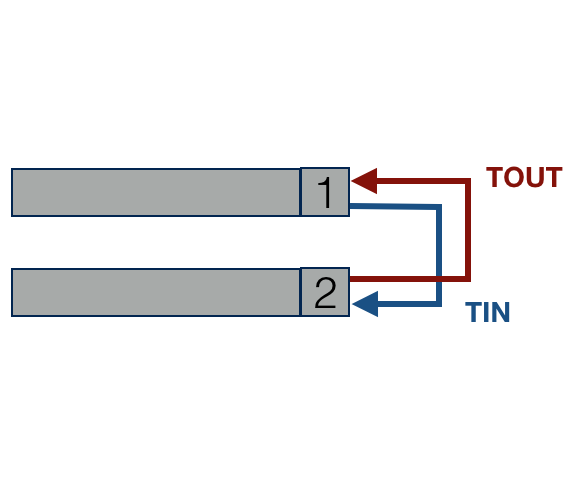}}
\caption[]{(a)~A schematic of the coincidence circuit in production. X and Y denote the modules (FEBs) used to identify X and Y positions. The red and blue lines represent the $50~\Omega$ coaxial cables looping around the plane. (b)~A schematic of the circuit used in the test; modules were placed with scintillating strips parallel to each other. 1 and 2 denote the modules~(FEBs) used in the test. The red and blue lines represent the $50~\Omega$ coaxial cables connecting two modules.}
\label{fig:triggers}
\end{figure}

The FEB trigger logic test was intended to examine the trigger logic generation of each FEB to generate the cross-module coincidences. Differences between module configurations used for generating triggers in production and in this test are shown in figure~\ref{fig:triggers}. In the test, two CRT modules placed on the wooden A-frame with parallel strip directions were used to test all FEBs. Two event rates of one module with and without TIN inputs were compared. The event rate of crossing cosmic muons of an isolated module is expected to drop when requiring the trigger signal from another module. The test showed that any FEB combination resulted in a significant event rate drop with respect to that of a single isolated FEB.


\subsection{Module tests}

Each CRT module is composed of 16 scintillating strips each with two wavelength-shifting fibers and SiPMs enclosed in a protective aluminum case. The module works by detecting the strip's scintillation light, which indicates the need to block all external light sources within the module. After receiving modules at DAB, we performed a series of tests to examine the quality of modules as well as scintillating strips and the tagging efficiency of each module.

\subsubsection{Light leak}

\begin{figure}[htb!pb]
\centering
\includegraphics[width=0.55\textwidth]{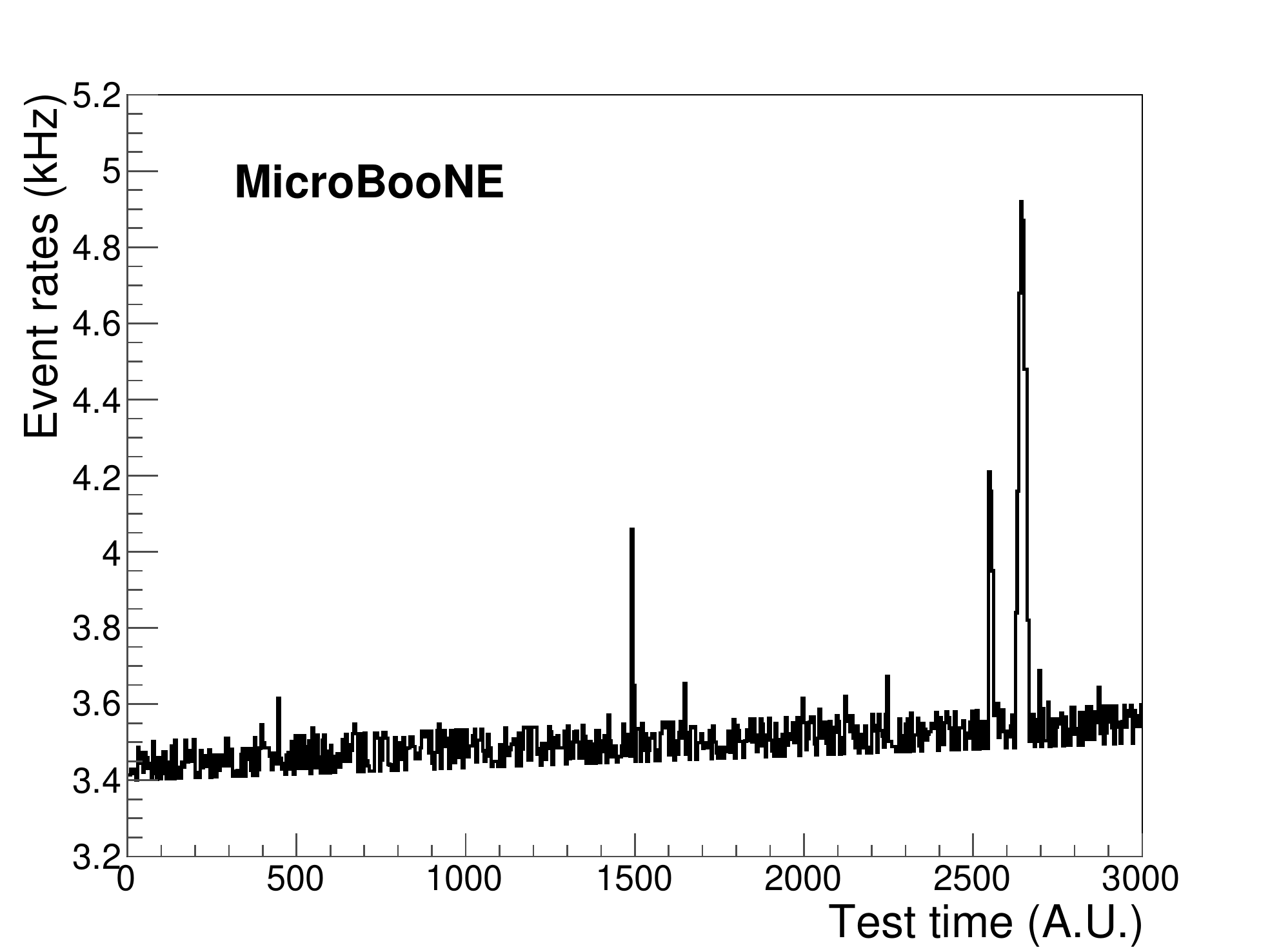}
\caption[]{Event rate of one module during light leak test. The rate remained stable until a light leak was found. At time counts around 2600, we moved the light source back and forth to locate where the light leak is. A glitch of event rates was found at around time counts 1500.}
\label{fig:lightleak}
\end{figure}

The light leak test was intended to detect any light leaks on the scintillating module that may have occurred during shipping. In the test, a commercial 80-watt lamp~(producing 1200 lm) was used to illuminate the modules. The event rate of the tested module was recorded as the light source was moved at a consistent speed along the module. As seen in figure~\ref{fig:lightleak}, the event rate remained stable until encountering a light leak. We fixed light leaks by patching the enclosure with opaque Epoxy glue and we subsequently re-tested the module. 

\subsubsection{Scintillating strip response}
\label{sec:subsub_channel_test}

\begin{figure}[htb!pb]
\centering
\subfloat[Channels 1 and 2 responses to source.]{\label{fig:source_test_a}\includegraphics[width=0.495\textwidth]{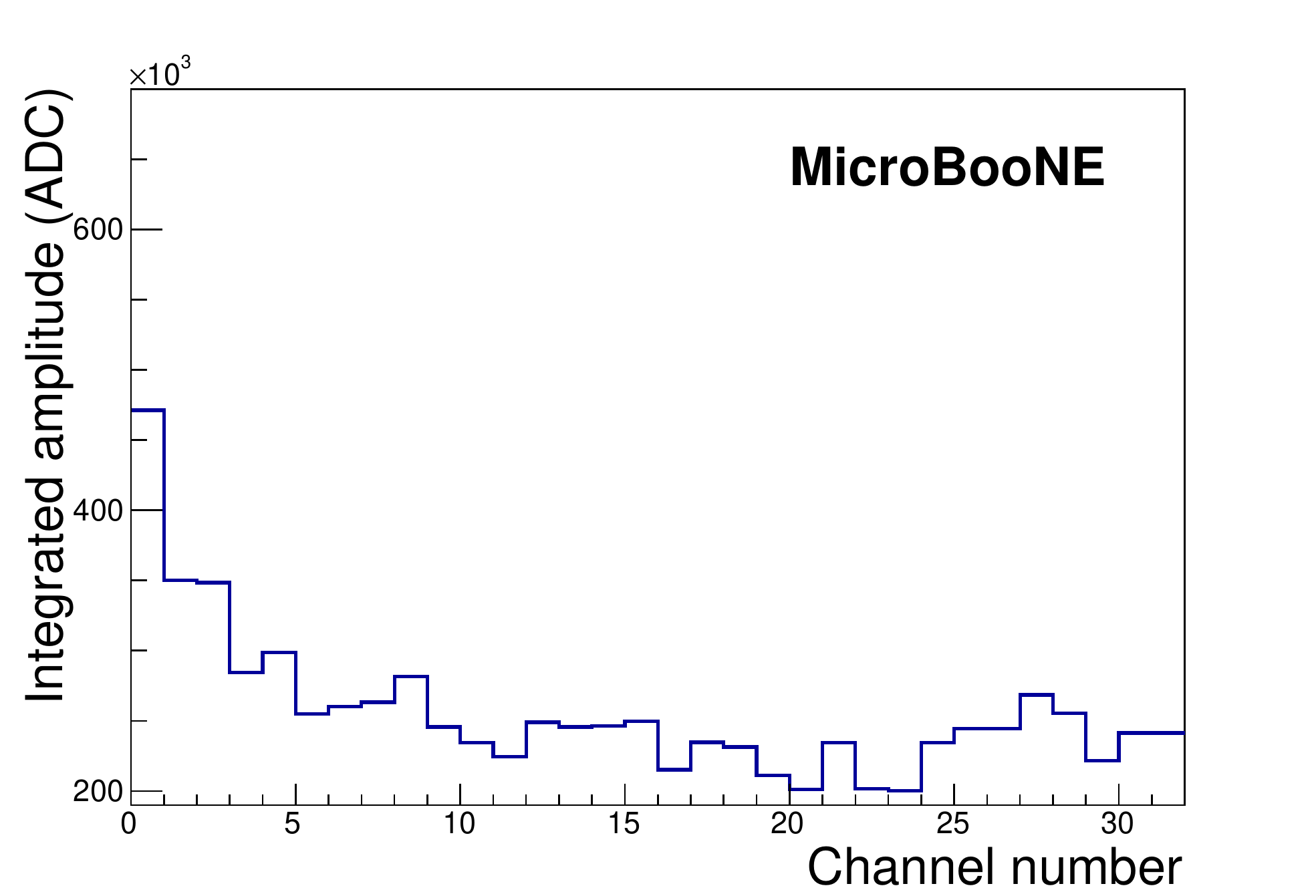}}
\hspace{\fill}
\subfloat[All channels responses to source.]{\label{fig:source_test_b}\includegraphics[width=0.495\textwidth]{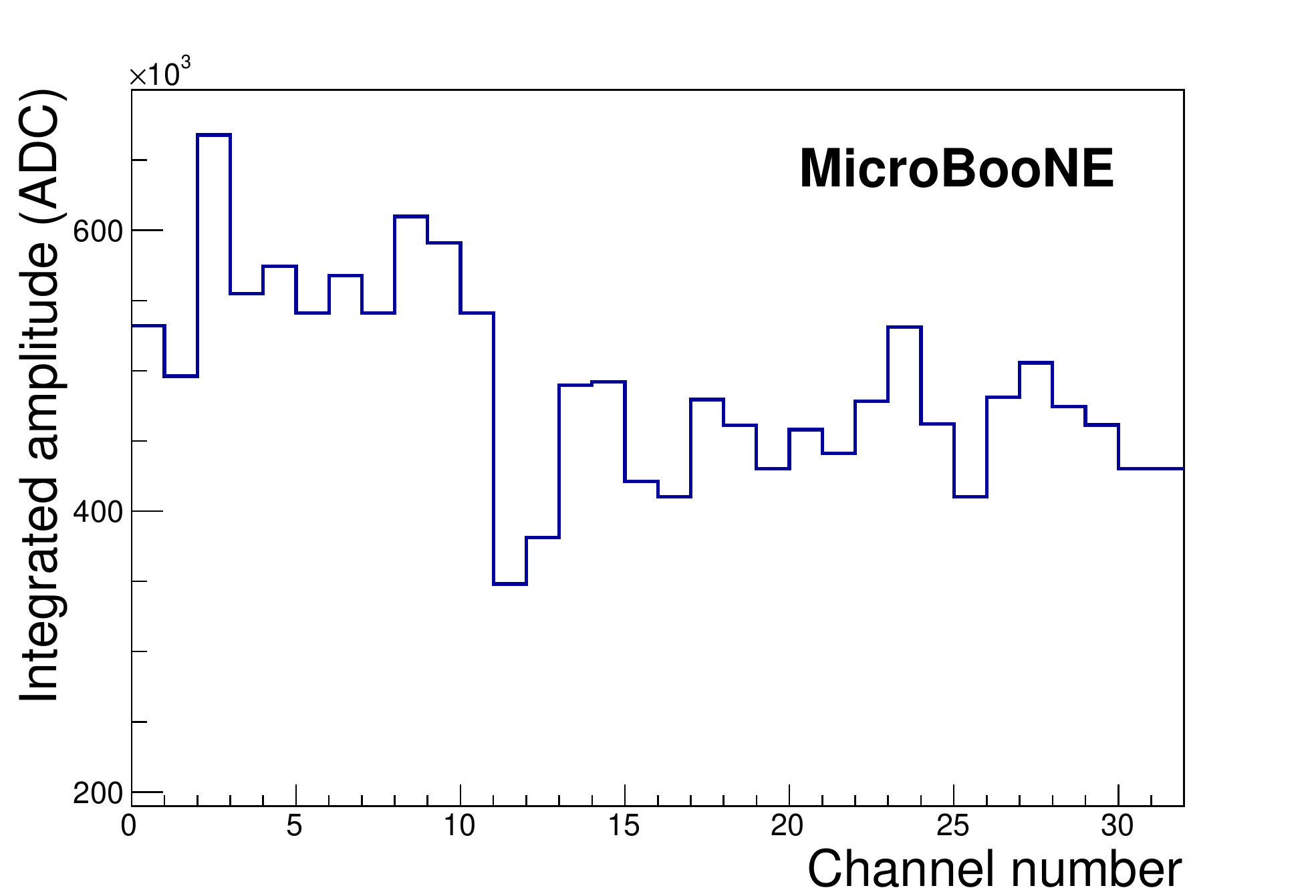}}
\caption[]{(a)~Example of high accumulated ADC counts on channel 0 in the module as the radioactive source approaches. (b)~Responses of 32 channels to the source. Channel 11 is an example of a broken channel, where only noise is detected. The low ADC counts accumulated on channel 12 results from a high threshold set for that channel which has been properly tuned.}
\label{fig:source_test}
\end{figure}

We conducted a radioactive source test to check the presence of any broken scintillating strips or fibers inside the modules. For this purpose, we used one radioactive $^{60}$Co gamma source. The source used at DAB and LArTF had an activity of about 74 kBq. During the test, the accumulated analog to digital converter~(ADC) counts on each channel was monitored as the source was moved transversely across all scintillating strips slowly. A fully-functioning channel had a positive response to the source as shown in figure~\subref*{fig:source_test_a}. An example of a failed strip found is shown in figure~\subref*{fig:source_test_b}. Only 12 channels were found to be broken out of the total 2,336 channels after the test. The connectors of broken channels were shorted inside the FEB. No data is taken on these channels during MicroBooNE physics data-taking. The test was repeated during commissioning as described in section~\ref{chapter:commissioning}.


\subsubsection{Muon tagging efficiency}

To further access damage to modules, we tested their ability to perform moon tagging. This was a complete test of both the module as well as its FEB readout and triggering. All modules were tested in groups of three. In this test, we constructed two sets of telescopes, a two-layer muon telescope made of two modules excluding the tested module, and a three-layer muon telescope of the three modules in one group.



\begin{figure}[htb!pb]
\centering
\includegraphics[width=0.6\textwidth]{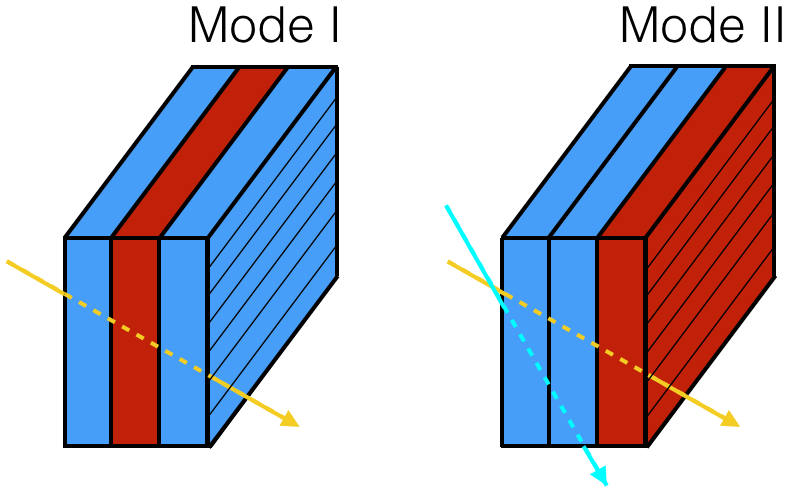}
\caption[]{In Mode I, two-layer and three-layer muon telescopes were triggered by the same flux of cosmic muons~(yellow). In Mode II, the two-layer muon telescope was triggered with additional cosmic muons~(cyan).}
\label{fig:cosmiccarton}
\end{figure}

Muon tagging efficiency is defined as {$N_3$/$N_2$}, where $N_2$ is the number of cosmic muons triggering the two-layer telescope and $N_3$ is the number of cosmic muons triggering the three-layer telescope. $N_3$ was recorded when $N_2$ reached 10,000 total triggers. Figure~\ref{fig:cosmiccarton} shows the schematic of two modes of muon telescopes used in this test. In Mode I, the tested module was located between the two-layer muon telescope. In Mode II, the tested module was the outer module. Mode II was necessary due to handling complications at DAB.

\begin{figure}[htb!pb]
\centering
\subfloat[Cosmic tagging efficiencies.]{\label{fig:efficiency-a}\includegraphics[width=0.495\textwidth]{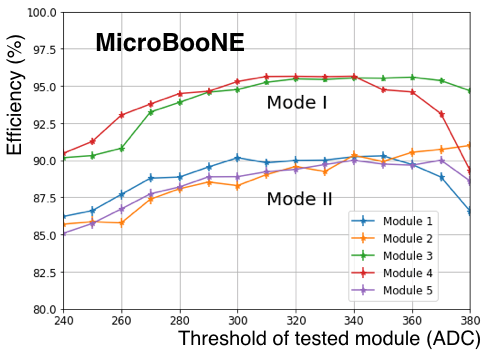}}
\hspace{\fill}
\subfloat[Event rates of tested modules.]{\label{fig:efficiency-b}\includegraphics[width=0.478\textwidth]{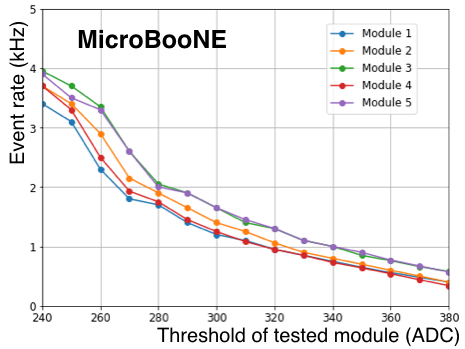}}
\caption[]{(a)~Efficiencies of five modules with Modules 3, 4 tested in Mode I and Modules 1, 2, 5 tested in Mode-II. The background noise has a significant impact on the muon tagging efficiency before the module threshold value is high enough ($\sim$290 ADC counts). In the region of 290-340 ADC counts, the background noise is reduced. Tagging efficiencies are $\sim$95\% for Mode I and $\sim$90\% for Mode II. Efficiencies in Mode II are lower due to muons that pass the first two modules at large angles miss the third one. (b)~The event rate has a linear correlation to the threshold value in the same region. Error bars show statistical uncertainties only.}
\label{fig:efficiency}
\end{figure}

Muon tagging efficiencies were tested with different ADC threshold values from 240 ADC counts to 380 ADC counts~(a conversion gain test~\cite{crt_novel} indicated one photoelectron has $\sim$290 ADC counts and two photoelectrons have $\sim$350 ADC counts). As shown in figure~\subref*{fig:efficiency-a}, the efficiencies in Mode I are higher than the efficiencies in Mode II due to a larger flux of detectable cosmic muons in Mode II. In Mode II, muons that pass the first two modules at large angles  miss the third one. As the threshold increases, the muon tagging efficiencies increase compared to the low threshold value region as the noise interference, which is uncorrelated between modules, decreases. At larger threshold, signals produced by cosmic muon-induced light are rejected, which results in a lower tagging efficiency. Trigger rates for individual modules decrease as module ADC threshold increases, as shown in figure~\subref*{fig:efficiency-b}. Event rates show a linear correlation with ADC threshold after the threshold value is large enough to remove background noise.

After being installed at LArTF, CRT planes are separated by several meters~(modules were placed more closely in the test). Ref.~\cite{crt_novel} has demonstrated that the CRT maintains the same excellent timing resolution at this distance. Hence, in MicroBooNE's CRT, with proper strip-to-strip calibration (not implemented in these tests), we expect to achieve a higher (>95\%) level of cosmic muon tagging efficiency.

\subsection{Commissioning tests}
\label{chapter:commissioning}
During commissioning, we tested the Wiener PL512 outputs and GPS timestamps/beam timing signals. The sensing voltage on each output of the Wiener PL512 was checked with FEBs powered up one-by-one on each line. The amplitudes of GPS timestamps and beam timing signals were found at the desired values at the end of each line using an oscilloscope.

The channel response test described in section~\ref{sec:subsub_channel_test} was repeated for each module after the full installation to examine potential damage caused during installation. Only two additional scintillating strips were found to be non-operative. 

Following the full testing regime described above, a total of 14 CRT channels out of 2,336 were identified as unusable. This means that about 0.6\% of the CRT module surface will have a position resolution in one direction that is reduced from the cm-level to the width of a strip, 10.8~cm~\cite{crt_novel}, without a calibration with two WLS. This is expected to have a negligible impact on the physics delivered by the CRT sub-system.

\section{Construction and integration at LArTF}
\label{sec:reception}

CRT Modules and FEBs were shipped from the University of Bern to Fermilab in several batches. Eleven frames were used to ship 73 modules; shipping details are summarized in table~\ref{T:frames}. The following sections describe CRT receiving, construction layouts and procedures.

\begin{table}[htb!pb]
\caption{73 modules were built and shipped to Fermilab between May 2016 and December 2016.}
\begin{tabularx}{\textwidth}{ c *{12}{Y|} }
\hline
\multicolumn{1}{| c |}{\textbf{Time}} & 
\multicolumn{3}{ c |}{\textbf{May 2016}} & 
\multicolumn{5}{ c |}{\textbf{May-July 2016}} & 
\multicolumn{3}{ c |}{\textbf{December 2016}} & 
\multicolumn{1}{ c |}{\textbf{Total}} \\ 
\hline
\multicolumn{1}{| c |}{Frame} &1 & 2 &3 &4 &5 &6 &7 &8 &9 &10 &11  & 11\\ 
\hline
\multicolumn{1}{| c |}{Modules (no.)} & 8  & 6 & 7 & 6 & 6 & 6 & 5 & 5 & 6 & 7 & 11  & 73\\ 
\hline
\label{T:frames}
\end{tabularx}
\end{table}


\subsection{Receiving and pre-installation test}
During shipping, modules were placed on A-frames and were shrink-wrapped together, separated by safety strips and one additional layer of foam for further firmness and protection. All wooden surfaces in contact with the modules were also covered in thick foam. After receiving, modules were stored and tested in DAB at Fermilab as previously described in section~\ref{sec:test}. 

\subsection{Electronic racks}

\begin{figure}[htb!pb]
\centering
\includegraphics[width=0.6\textwidth]{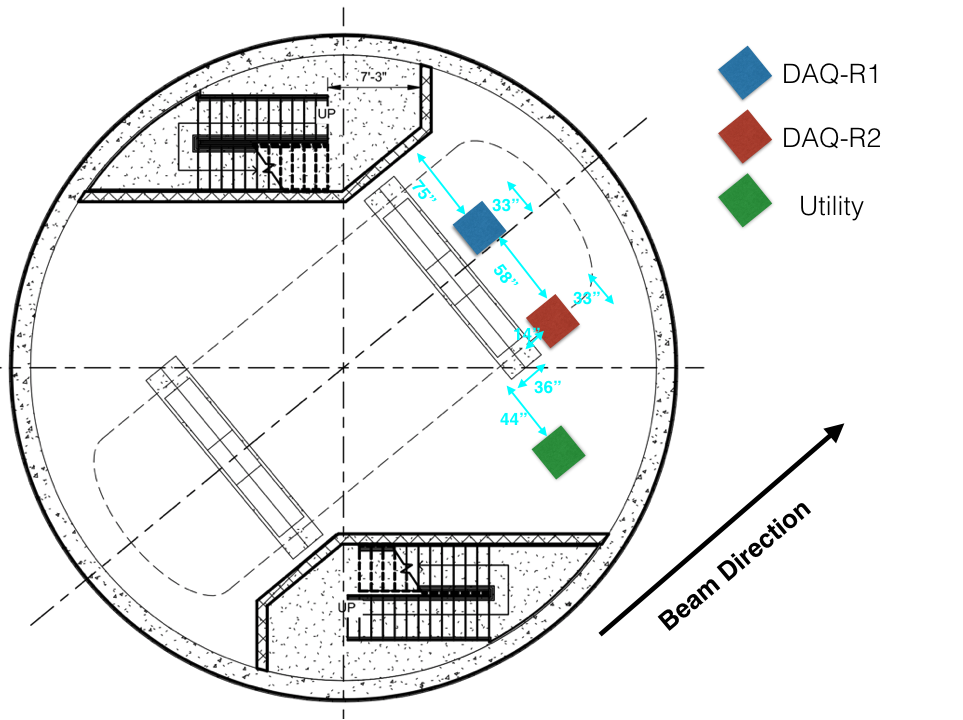}
\includegraphics[width=0.85\textwidth]{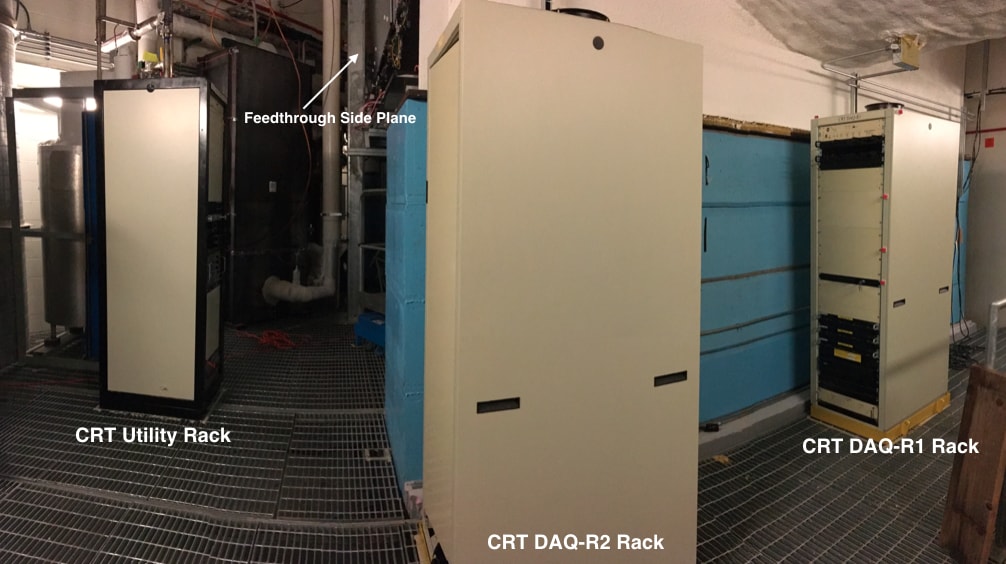}
\caption[]{CRT racks in the LArTF pit. Three racks are located in a free area at the downstream beam end under the cryostat and next to the feedthrough side plane.} 
\label{fig:pitracklocations}
\end{figure}

CRT electronic racks were fully constructed at DAB and then shipped to LArTF. During installation at LArTF, electronic racks were lowered to the pit area using the LArTF crane. The layout of electronic racks is shown in figure~\ref{fig:pitracklocations}. Rack components are described in section~\ref{sec:rack}.

\subsection{Module installation phase I}
Module installation phase I involves installation and integration of the bottom, pipe side, and feedthrough side planes. Modules, along with installed cables are summarized in table~\ref{tab:phaseatable}. 

\begin{table}[htb!pb]
\centering
\caption{Phase I installation and commissioning. The numbers in parentheses indicate channels found to be non-operative during testing and commissioning.}
\begin{adjustbox}{max width=\textwidth}
\begin{tabular}{|l|c|c|c|c|c|c|c|c|}
    \hline
    Plane & Date & Modules & FEBs & Channels & GPS Lines & Beam Timing Lines & DC Lines & Cables\\
    \hline
    Bottom		& Jul2016 & 9  & 9  & 144(0)  & 1 & 1 & 2 & 61\\
	\hline
	Feedthrough	& Sep2016 &	13 & 13	& 208(4)  & 1 & 1 &	2 &	85\\
	\hline
	Pipe		& Sep2016 &	27 & 27 & 432(10) &	4 &	4 &	3 &	171\\
    \hline
    \end{tabular}
\label{tab:phaseatable}
\end{adjustbox}
\end{table}

\subsubsection{Bottom plane installation}
\begin{figure}[htb!pb]
\centering
\includegraphics[width=0.85\textwidth]{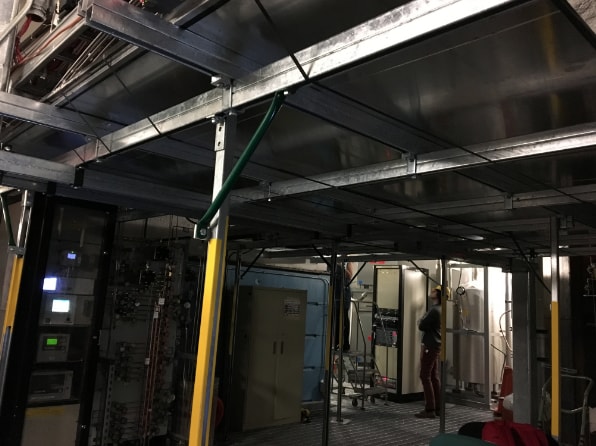}
\caption[]{MicroBooNE CRT showing the bottom plane and its support structure}
\label{fig:installedbottom}
\end{figure}

Bottom plane modules cover the area between the saddles that support the cryostat. The modules, along with their support structure, are located underneath the MicroBooNE cryostat in the pit of the LArTF cryostat, as seen in figure~\ref{fig:design}. There are two di-layers of modules in the bottom plane, one seven-module square and one smaller two-module square as summarized in table~\ref{T:modules}. The support structure is composed of a flat table to physically hold the plane and the support legs. The completed bottom plane is shown in figure~\ref{fig:installedbottom}.


The first step was building and assembling the support structure on the ground underneath the cryostat. 
It consists of 8 pieces of B22A unistrut~\cite{b22a} channels as the columns, and a set of B22A double-channels unistrut for the table top manifold. The assembly supports the set of bottom plane modules with a total mass of about 1,100 kg.

\begin{figure}[htb!pb]
\centering
\includegraphics[width=0.85\textwidth]{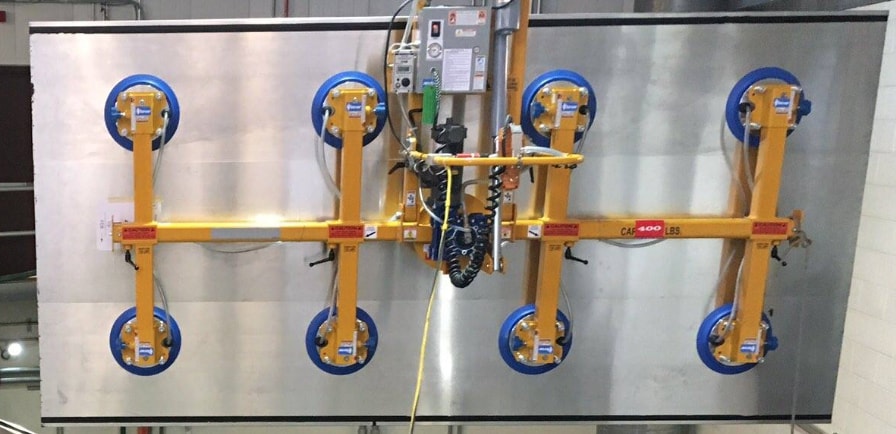}
\caption[]{Rigging system used to grip and move CRT modules.}
\label{fig:lifttingjig}
\end{figure}

After the support structure was constructed on the ground, modules were lowered down using the roof crane and a rigging system. The rigging system is equipped with a commercial vacuum lifting fixture designed and fabricated by ANVER\textsuperscript{\textregistered}~\cite{anver}~(figure~\ref{fig:lifttingjig}) and attached to the crane. The vacuum lifting fixture can be rotated to accommodate any installation orientation. The rigging system was used to lift modules onto the support structures. 
 
\subsubsection{Feedthrough side and pipe side plane installation}

The two side planes on the long sides of the MicroBooNE cryostat are supported with a curtain-system trolley assembly. This approach suits the situation at LArTF for two reasons:
\begin{itemize}
\item {There is very limited space between the MicroBooNE platform and LArTF building walls. This approach allows modules to be hung in the work area and rolled into tight places.}
\item {It yields a sturdy structure that is able to move at any time if access through a module is needed.}
\end{itemize}

There are five sets of trolley assemblies for the two side planes, two for the feedthrough side plane with one inner side assembly carrying seven modules and one outer side assembly carrying six modules, and three for the pipe side plane, carrying seven, ten and ten modules. 

\begin{figure}[htb!pb]
\centering
\subfloat[Trolley assembly for side planes.]{\label{fig:roller-a}\includegraphics[width=0.495\textwidth]{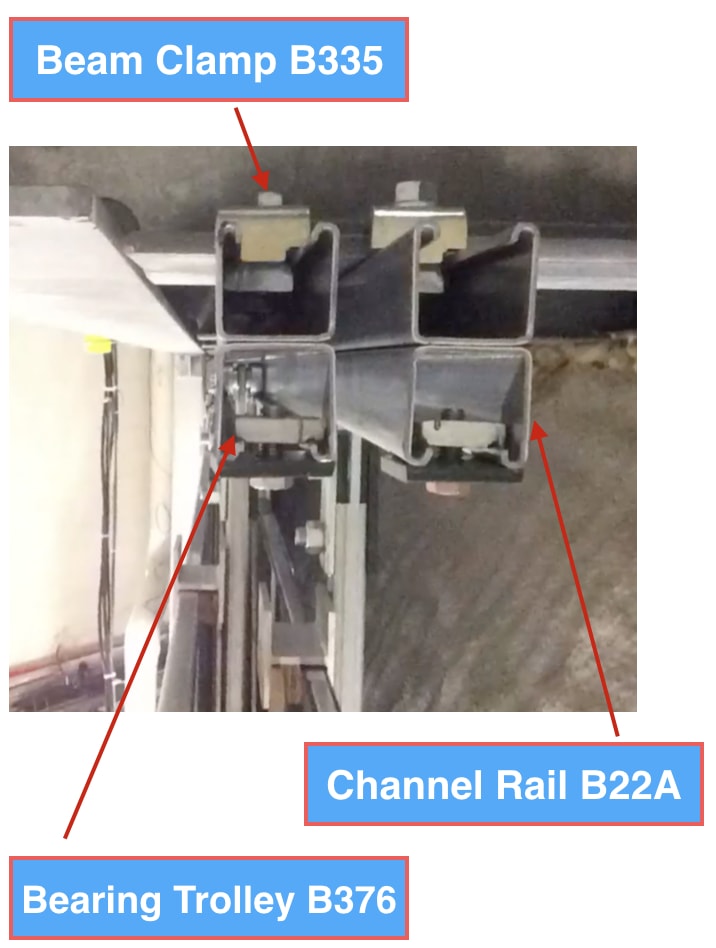}}
\hspace{\fill}
\subfloat[Installed module held by B line and H clip.]{\label{fig:roller-b}\includegraphics[width=0.478\textwidth]{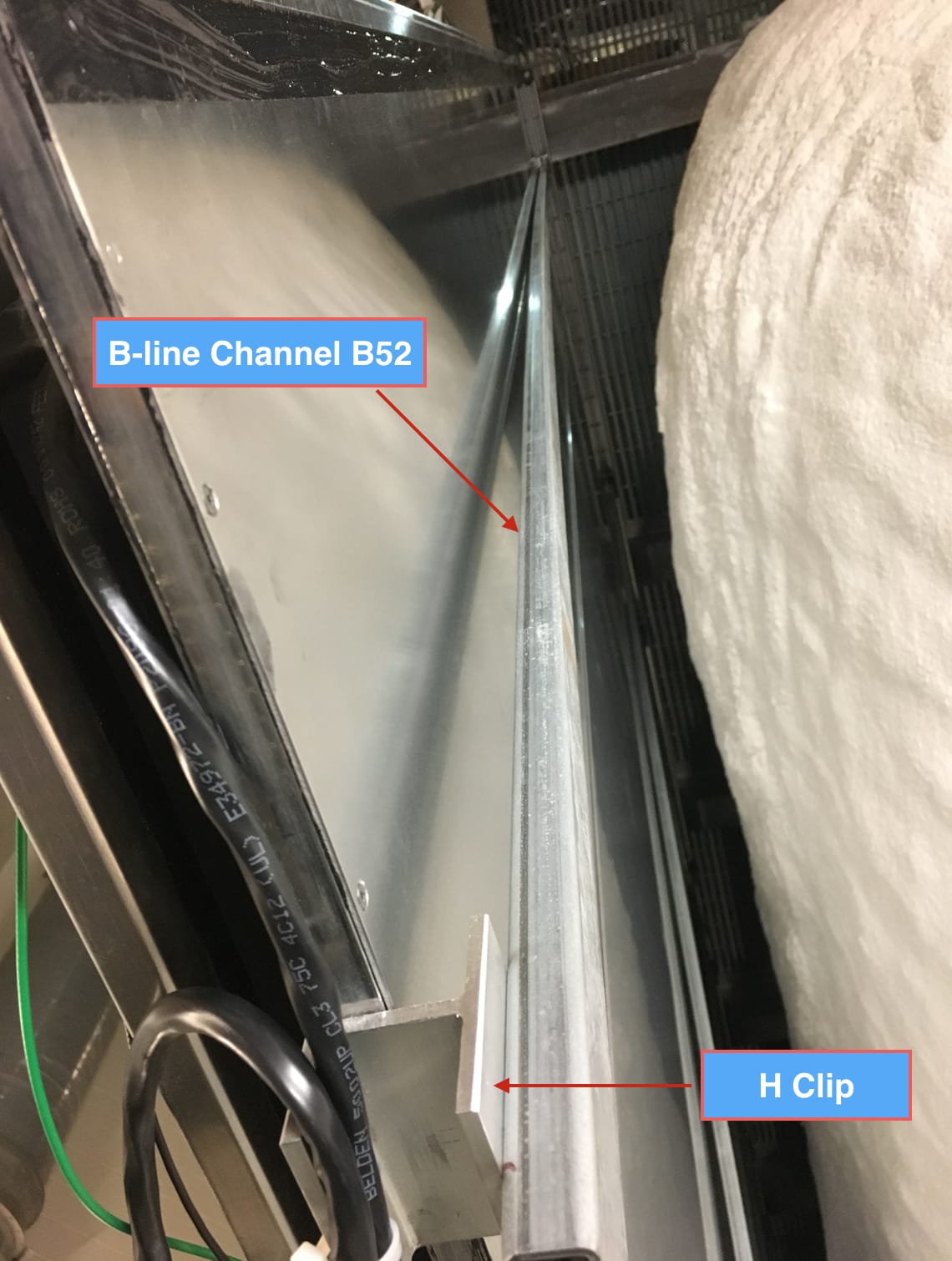}}
\caption[]{(a)~Profile of the trolley assembly installed at the top of the feedthrough side plane. (b)~One B-line channel B52 and one aluminum H clip used to hold modules on the feedthrough side plane.}
\label{fig:rollers}
\end{figure}

Each trolley assembly is made of a set of B-line bearing trolleys B376 \cite{b376} rolling along the B22A double-channel unistrut as shown in figure~\ref{fig:rollers}. These channels are clamped on the existing I-beams with the longest span about 2.7 m for the feedthrough side plane and 1.1 m for the pipe side plane. We use beam clamps B355, which have a load capacity of 540 kg each. As there are at least 6 I-beams in total and subsequently 12 beam clamps needed for each rail assembly, the total load capacity provided by the clamps is 6,500 kg, much higher than the mass of either side plane. 

\begin{figure}[htb!pb]
\center
\includegraphics[width=0.9\textwidth]{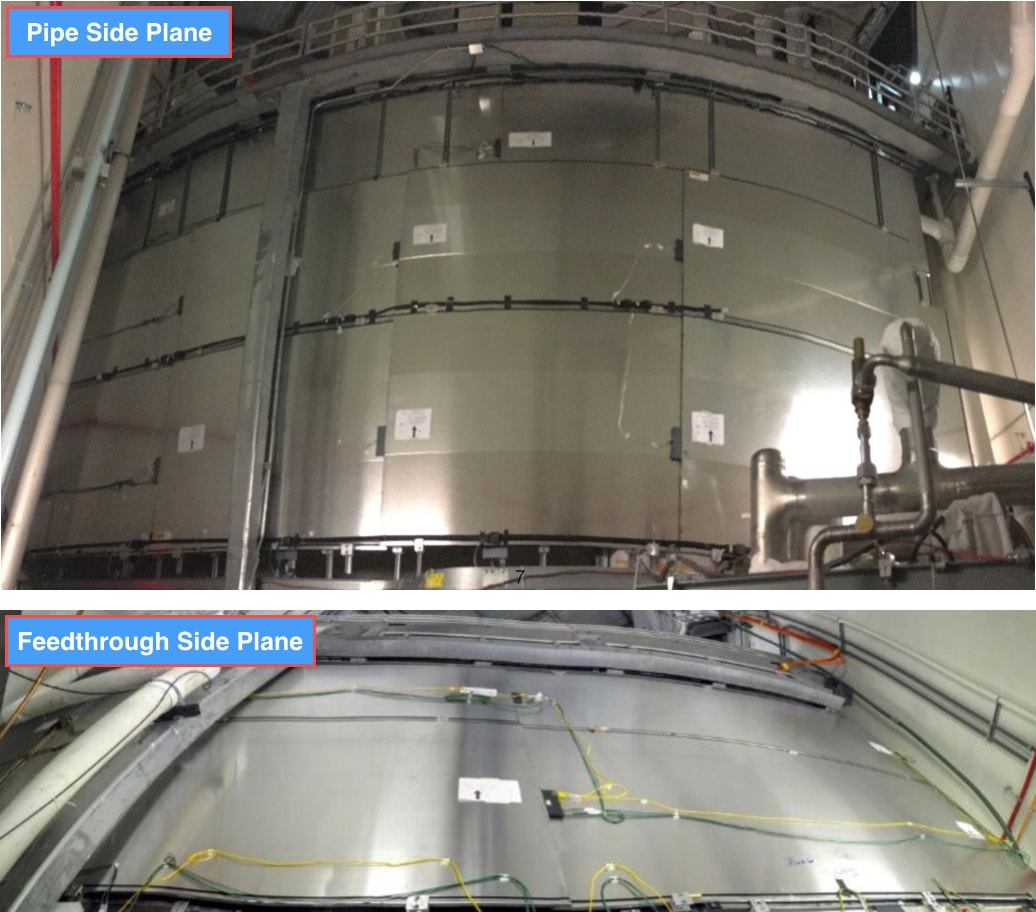}~
\caption{MicroBooNE CRT showing the pipe side plane~(picture distorted, taken from side) and feedthrough side plane~(picture take from bottom). }
\label{fig:pipeandft}
\end{figure}

One B-line channel B52 is hung from each bearing trolley. With proper fittings, these channels are used to cradle and support side CRT modules. Hex head cap screws are used for fastening parts to the bearing trolley assembly. Aluminum H clips connected to the B52 channel are used to hold the modules. 
Installed pipe side and feedthrough side planes are shown in figure~\ref{fig:pipeandft}.


\subsection{Module installation phase II}
Installation Phase II involves installation and integration of top plane modules as summarized in table~\ref{tab:phasebtable}. The top plane is installed above the MicroBooNE platform above the cryostat. The top module group, consisting of 16 long modules and 8 small modules with a total mass of 3,400 kg, is arranged in 2 layers and are supported by a table-top manifold with 14 legs. The top plane module configuration is documented in table~\ref{T:topmoduledetails}.

\begin{table}[htb!pb]
\centering
\caption{Phase II installation and commissioning summary. The numbers in parentheses indicate channels found non-operative during testing and commissioning.}
\begin{adjustbox}{max width=\textwidth}
\begin{tabular}{|l|c|c|c|c|c|c|c|r|}
    \hline
    Plane & Date & Modules & FEBs & Channels & GPS Lines & Beam Timing Lines & DC Lines & Cables\\
    \hline
	Top		& Feb2017 &	24 & 24 & 384(4) &	2 &	2 &	3 &	155\\
    \hline
\end{tabular}
\end{adjustbox}
\label{tab:phasebtable}
\end{table}

\begin{figure}[htb!pb]
\centering
\includegraphics[width=0.9\textwidth]{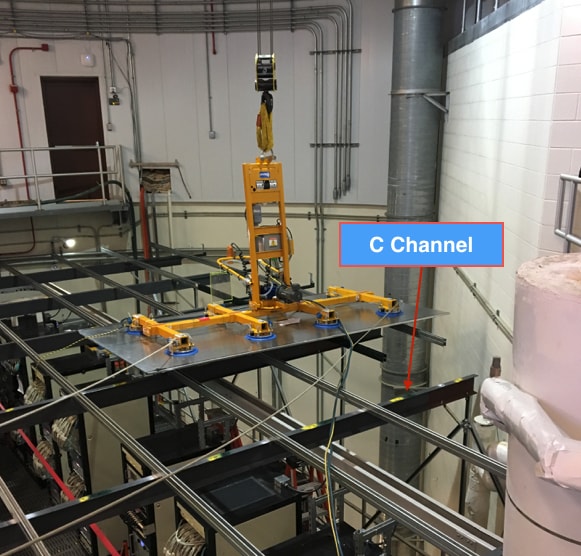}
\caption[]{MicroBooNE top plane support structure and C Channel. The roof crane and rigging system is placing one module in position.}
\label{fig:topcchannel}
\end{figure}

The support structure for the top plane, shown in figure~\ref{fig:topcchannel}, is made of unistrut provided by B-line, as with the bottom plane. In addition, six pieces of long C-channel span the space above the existing cabinets and cable trays on the MicroBooNE platform so that no additional support legs are needed in the platform walkway. Some C-channels were revised to extend out from the posts so that more channel supports could be added to minimize deflections. Additional B-line parts and braces were added so that a robust support table is formed. 

\begin{figure}[htb!pb]
\centering
\includegraphics[width=0.9\textwidth]{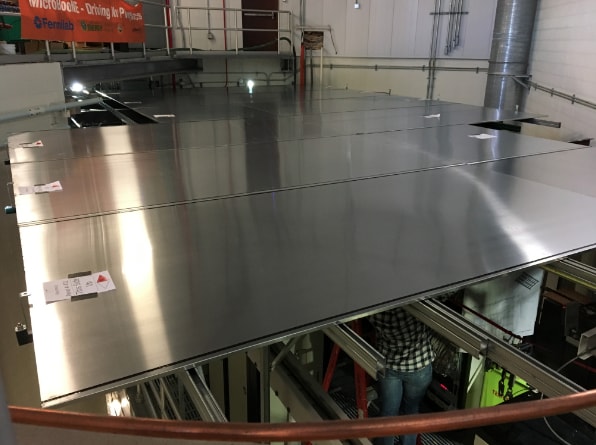}
\caption[]{Completed MicroBooNE CRT top plane.}
\label{fig:installedtop}
\end{figure}

B22 unistrut channels and post base items are used for the table legs, as with the bottom plane support system. About 2.8~m of vertical space separates the installed table top from the platform. When installed, the table top is clear of any existing hardware in this level of the building. Support leg bases are mounted to the existing I-beams with proper spacers so that no opening of the existing 3.8 cm fiberglass platform grating must be cut. The completed MicroBooNE top plane is shown in figure~\ref{fig:installedtop}.

\subsection{Cable integration}
\label{subsection:cableinstalltion}

There were three groups of cable installation: the DC power distribution, GPS and beam timing connections, and network connections. Cables for each plane were installed directly after the construction of that plane.


\subsubsection{DC power}
\label{sec:dcpower}

\begin{figure}[htb!pb]
\centering
\includegraphics[width=0.95\textwidth]{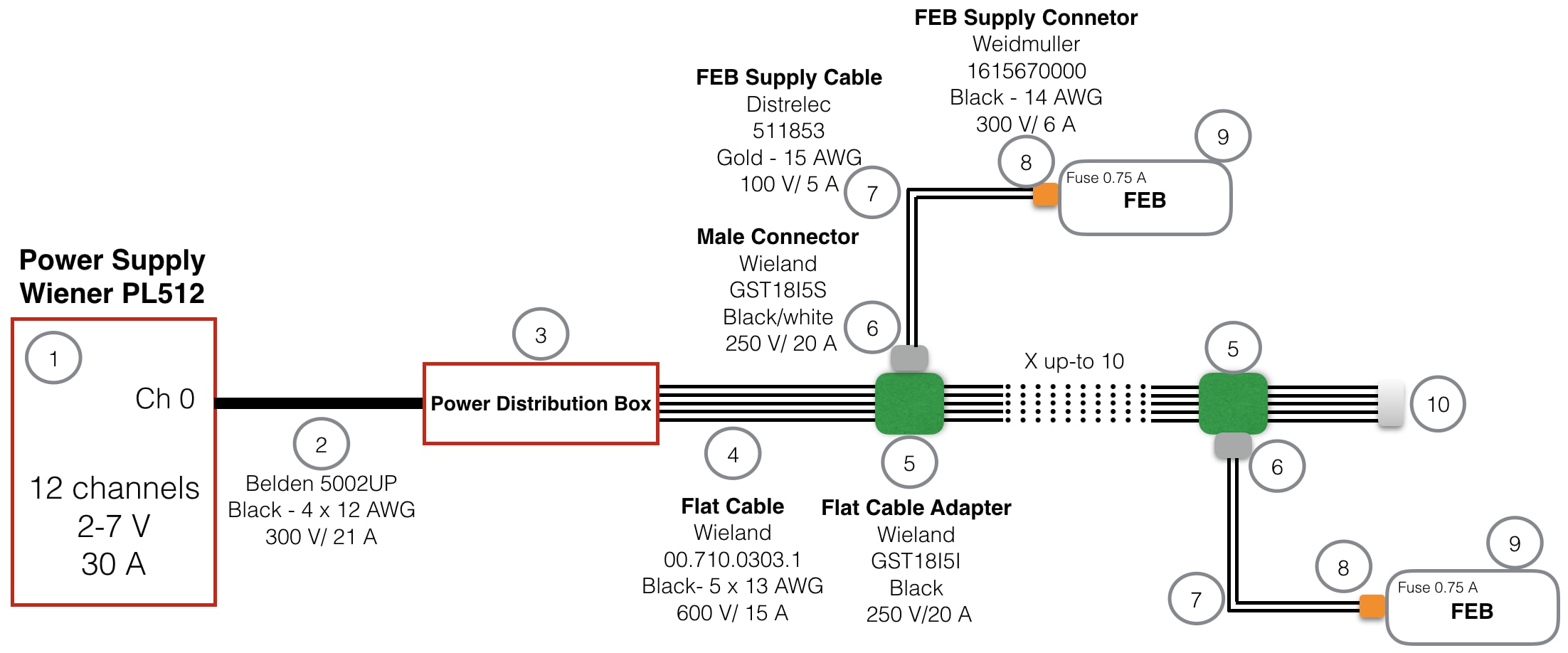}
\caption[]{Schematic of the CRT DC power distribution: (1)~Wiener power supply PL512 with 12 outputs of low voltage at 7~V maximum, (2)~Belden cable with left-end terminated with Andersen terminals, (3)~Customized power distribution box, (4)~Flat cable extended to reach up-to 10 FEBs, (5)~and (6), flat cable adapter and its male connector suitable for the (7)~FEB supply cable and, (8)~FEB connector. }
\label{fig:dcpowerschematic}
\end{figure}

The CRT DC Power Distribution is responsible for providing DC power to the 73 CRT FEBs. A schematic of the DC power distribution is shown in figure~\ref{fig:dcpowerschematic}. It consists of three parts: the Wiener PL512 low voltage power supply, the CRT power distribution box,  and a set of flat cable adapters connecting the FEB. 

As described in section~\ref{sec:rack}, a 12-channel commercial power supply Wiener PL512 is the source of low voltage (7V maximum) delivered to CRT power distribution boxes. The unshielded cable connecting the Powerpole housings on the back panel of the Wiener chassis to a CRT power distribution box contains four 12 AWG conductors. There are four wires in the cable with two wires for the power distribution box and the other two wires used as sense wire connections. The wires at one end of this cable are terminated with Andersen Powerpole terminals, enclosed in a housing and strain relief that mates with the panel mounted housing. The wires at the other end of the cable are terminated in ferrules that are inserted and secured into a Phoenix contact power distribution box \cite{phoenix} containing compression type contacts.

\begin{figure}[htb!pb]
\centering
\includegraphics[width=0.95\textwidth]{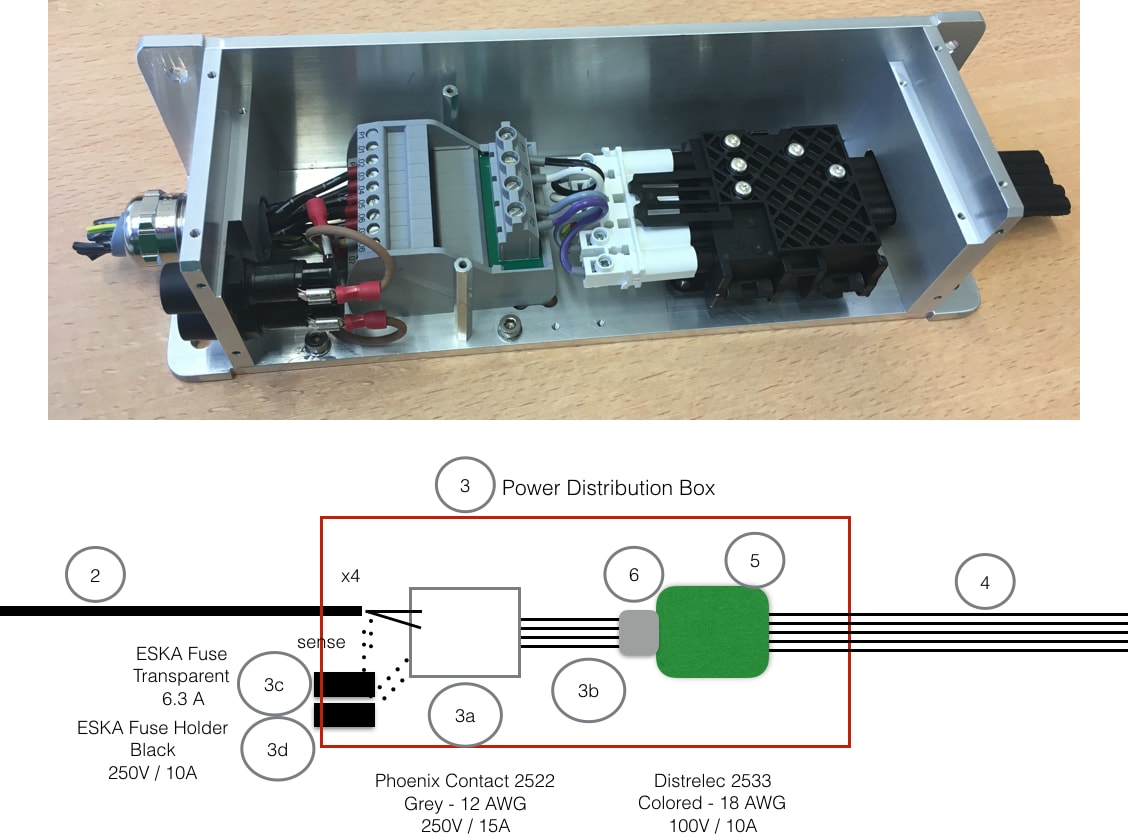}
\caption[]{Schematic of the CRT power distribution box: (3a)~Phoenix Connector 2522, (3b)~Flat cable, (3c)~6.3 A ESKA Fuse, (3d) ESKA Fuse Holder, (4)~Flat cable, (5)~and (6), flat cable adapter and its male connector.}
\label{fig:powerdistributionbox}
\end{figure}

The CRT power distribution box, shown in figure~\ref{fig:powerdistributionbox}, consists of power sensing fuses and the Phoenix contact power distribution block inside a protective aluminum box. Inside the box, the power is distributed into a Flat Cable \cite{flatcable5g2} which can be extended to power up to ten FEBs along the line.

\begin{figure}[htb!pb]
\centering
\includegraphics[width=0.95\textwidth]{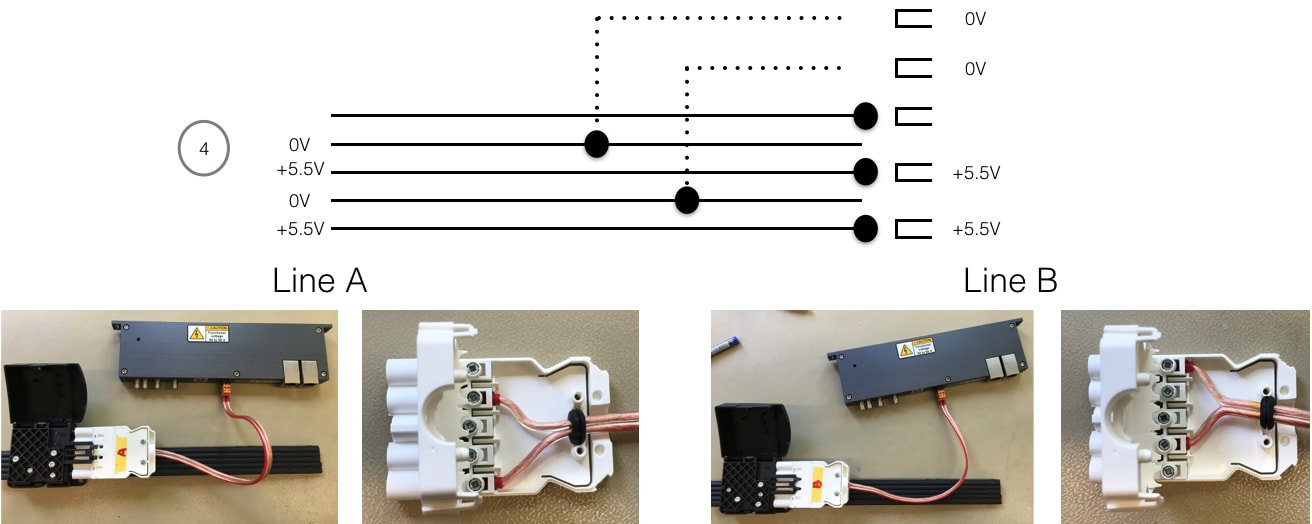}
\caption[]{Connection between the flat cable and FEB. Cable adapters are installed on the cable for two modes A and B. Corresponding male connectors are applied to match the FEB audio cable.}
\label{fig:maleconnectors}
\end{figure}

Along the flat cable, a few sets of flat cable adapter \cite{flatcableadapter} and male connectors \cite{flatcablemale} are installed to deliver power to each FEB as shown in figure~\ref{fig:maleconnectors}. Audio cable \cite{febcable} terminated with a specific pluggable terminal block \cite{febconnector} is used to connect the male connector and FEB power input.

\subsubsection{GPS and beam timing}
\label{sec:gpsandbeamtiming}
\begin{figure}[htb!pb]
\centering
\includegraphics[width=0.95\textwidth]{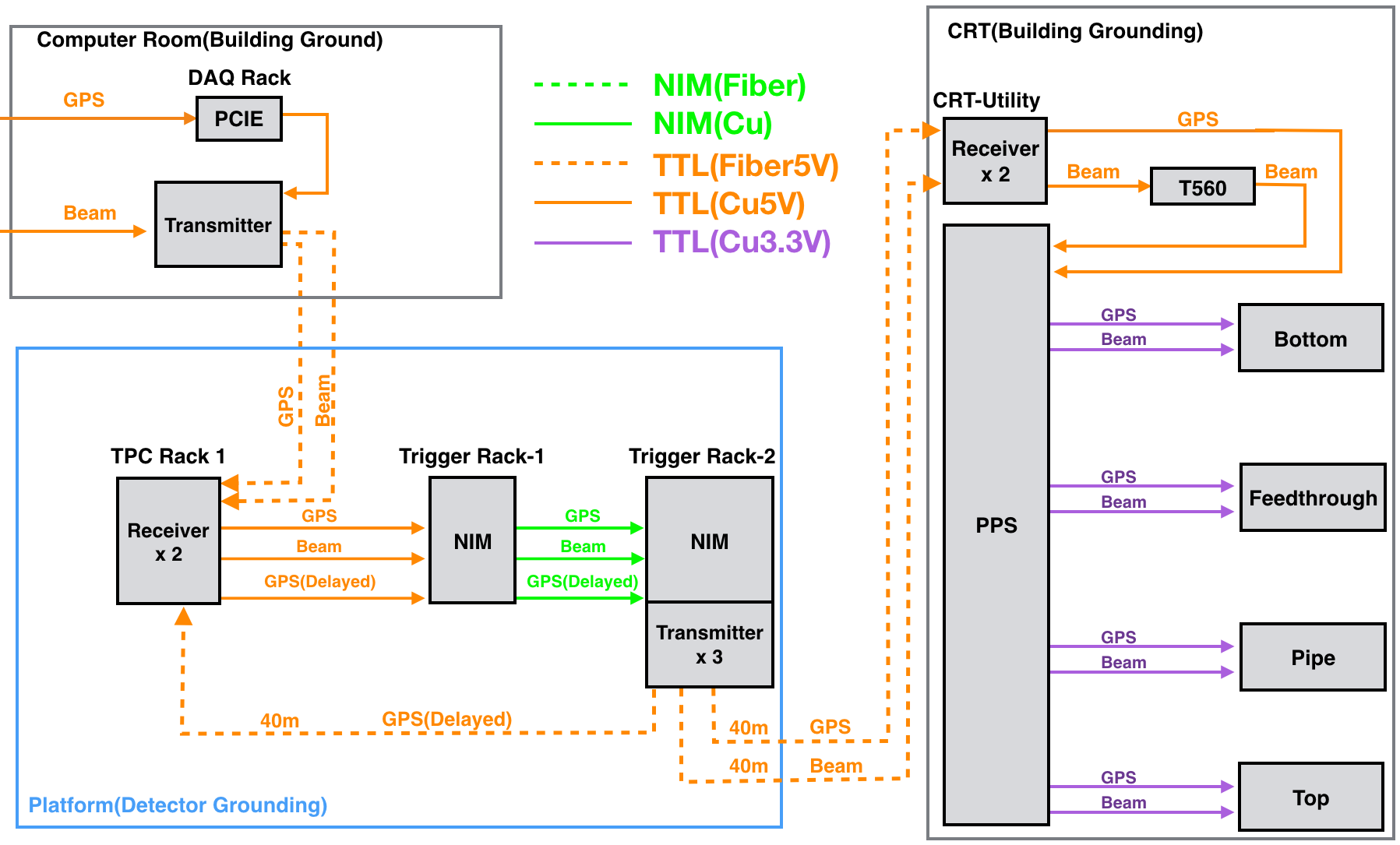}
\caption[]{GPS and beam timing distributions for the MicroBooNE CRT system at LArTF. On the platform, a third 40~m cable is used to generate an equally delayed GPS signal as the one delivered to the CRT building ground to compensate for the delay caused by long fiber cable transmissions between the platform and the CRT building ground. On the CRT building ground, the beam timing pulse is delayed for 40 $\mu$s by a delay unit (HighLand technologies, T560 4-channel compact digital delay and pulse generator~\cite{t560}) to prevent losing events due to FEB latency. }
\label{fig:timing.png}
\end{figure}

GPS timestamps and neutrino beam timing signals are crucial for the data matching between the MicroBooNE LArTPC and CRT. Transmissions of the two timing signals between MicroBooNE's private network and the CRT components are shown in figure~\ref{fig:timing.png}. 

GPS and beam timing signals are first received in the DAQ room at LArTF. The GPS timestamps come from an antenna on the building roof and are output through a PCIE card in a server in one of the MicroBooNE DAQ racks. Beam timing is received from the Accelerator Controls System. To avoid grounding issues, the two timing signals are delivered to MicroBooNE servers on the platform with a copper-fiber-copper transmission before subsequent re-routing to the CRT.

To further avoid grounding issues between LArTPC DAQ and CRT racks on the LArTF platform, three new signal transmitters (LuxLink DT-7201 \cite{transmitter}) and one new signal receiver (LuxLink DR-7201 \cite{transmitter}) are installed. Two of the transmitters comprise the copper-fiber transmissions for both timing signals to be delivered to the CRT racks. Two 40~m fiber cables are extended to connect the platform racks and CRT racks. The third transmitter and the receiver generate an equally delayed GPS signal as the one delivered to the CRT, with this delayed GPS timestamp incorporated into the normal MicroBooNE trigger board. In this way, the timing delay between the CRT and LArTPC data caused by long fiber cable transmission is mitigated. Two corresponding receivers are installed in the CRT Utility rack to receive both time signals. The beam timing pulse is additionally delayed for 40 $\mu$s by a high-precision delay unit (HighLand technologies, T560). This delay avoids losing events in the first 22 $\mu$s after the pulse due to FEB latency. With a customized component PPS designed by the LHEP at the University of Bern, timing signals are fanned-out to all CRT FEBs.

\subsubsection{Network}
\label{sec:network}
\begin{figure}[htb!pb]
\centering
\includegraphics[width=0.95\textwidth]{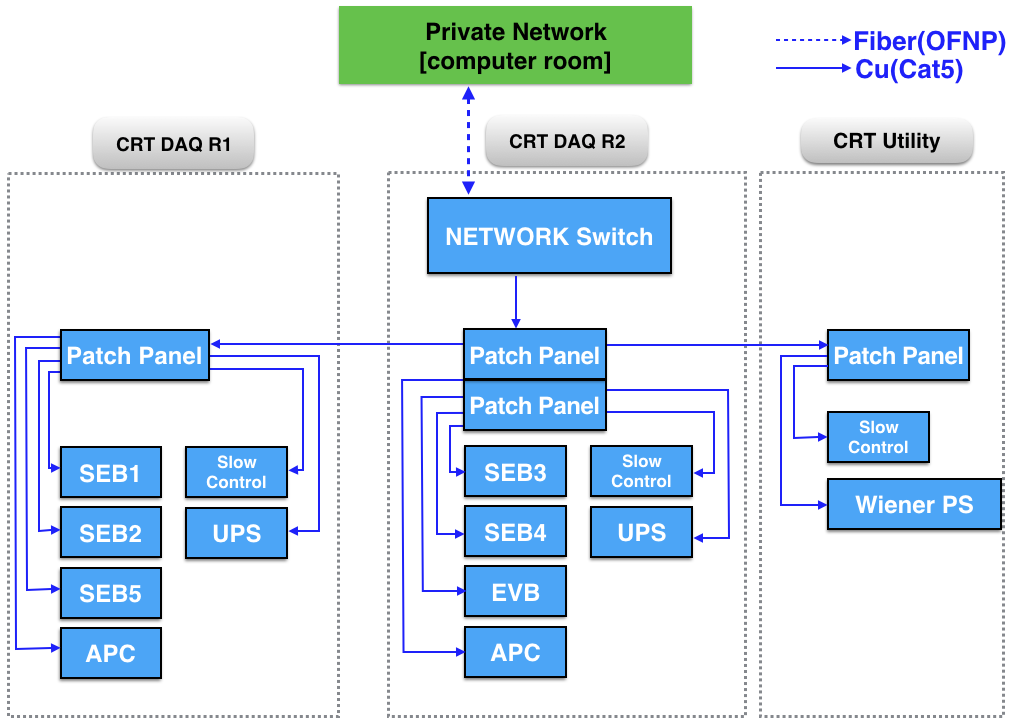}
\caption[]{Network and Data connections for MicroBooNE CRT system at LArTF.}
\label{fig:datanet.png}
\end{figure}

The CRT network connection incorporates CRT rack components and FEBs to the existing MicroBooNE private network. A physical network connection flow chart is shown in figure~\ref{fig:datanet.png}. The MicroBooNE private network originates in the MicroBooNE DAQ room and is extended to the network switch in the CRT's DAQ-R2 Rack by optical fiber 
connection. The network switch in DAQ-R2 plus patch panels in the other two CRT racks enable network connections to all CRT components. The FEBs (not shown) of each plane are daisy-chained to their closest servers using commercial Category 5 cable.

\section{CRT Operation and Performance}
\label{sec:performance}

MicroBooNE began data taking with the bottom and side planes in October 2016, and with the top plane in March 2017. In this section we present the operational status and performance of the MicroBooNE CRT.


\subsection{Operation status}

\begin{figure}[htb!pb]
\centering
\includegraphics[width=0.6\textwidth]{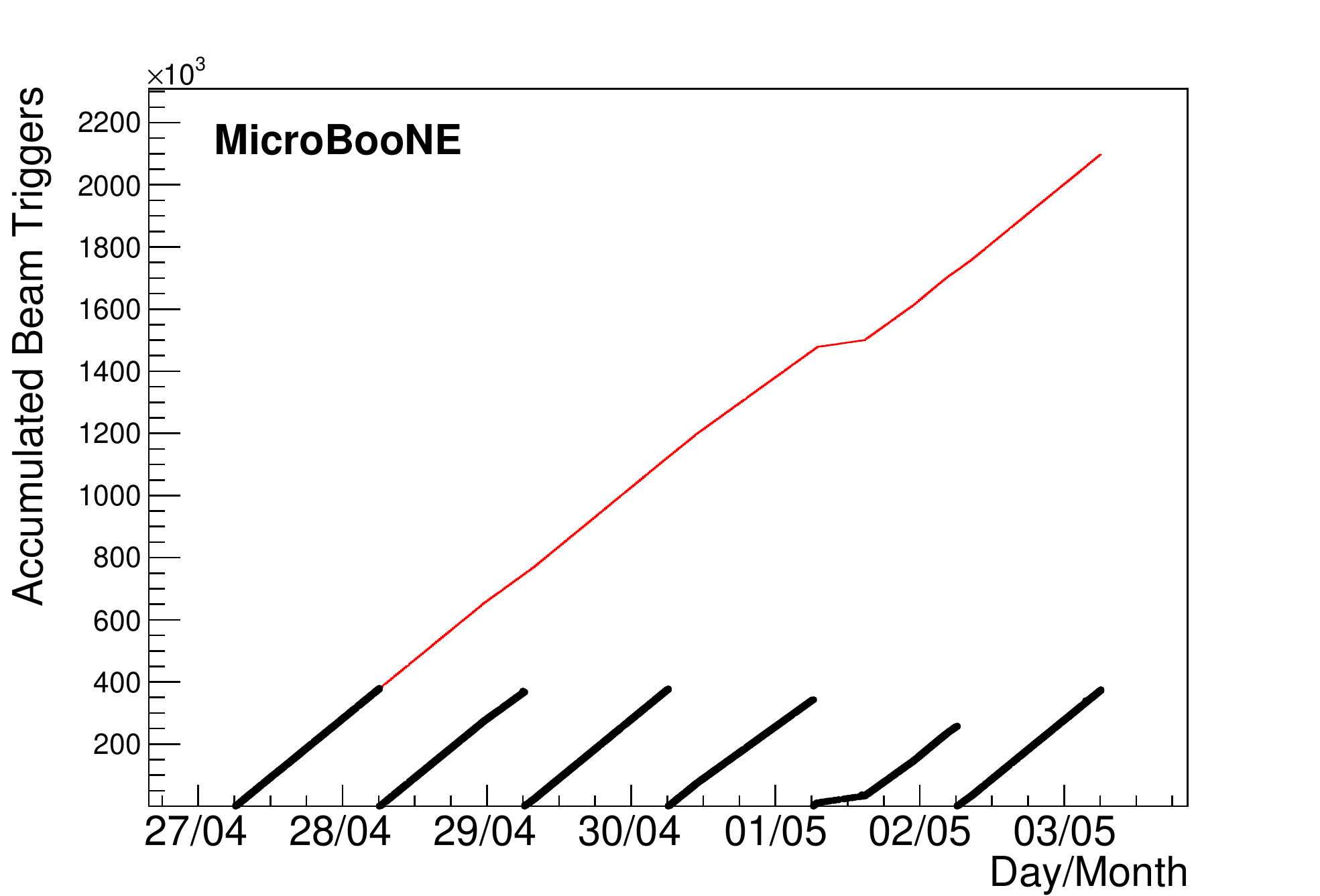}
\caption[]{The accumulated CRT event rates in the form of received beam triggers over one week. Black lines represent triggers received on each day. Red line represents the total triggers accumulated in the week. Beam was off for six hours on May 2nd. }
\label{fig:accumulated}
\end{figure}

When taking physics data, CRT events are self-triggered and use BNB triggers as a timing reference forwarded from the LArTPC DAQ racks. Beam triggers of BNB and neutrino at the main injector~(NuMI)~\cite{numi} provide the interaction times. 
Figure~\ref{fig:accumulated} is an example of daily accumulated CRT triggers.

\subsection{Neutrino events}
\label{sec:crtevents}

\begin{figure}[htb!pb]
\centering
\includegraphics[width=0.85\textwidth]{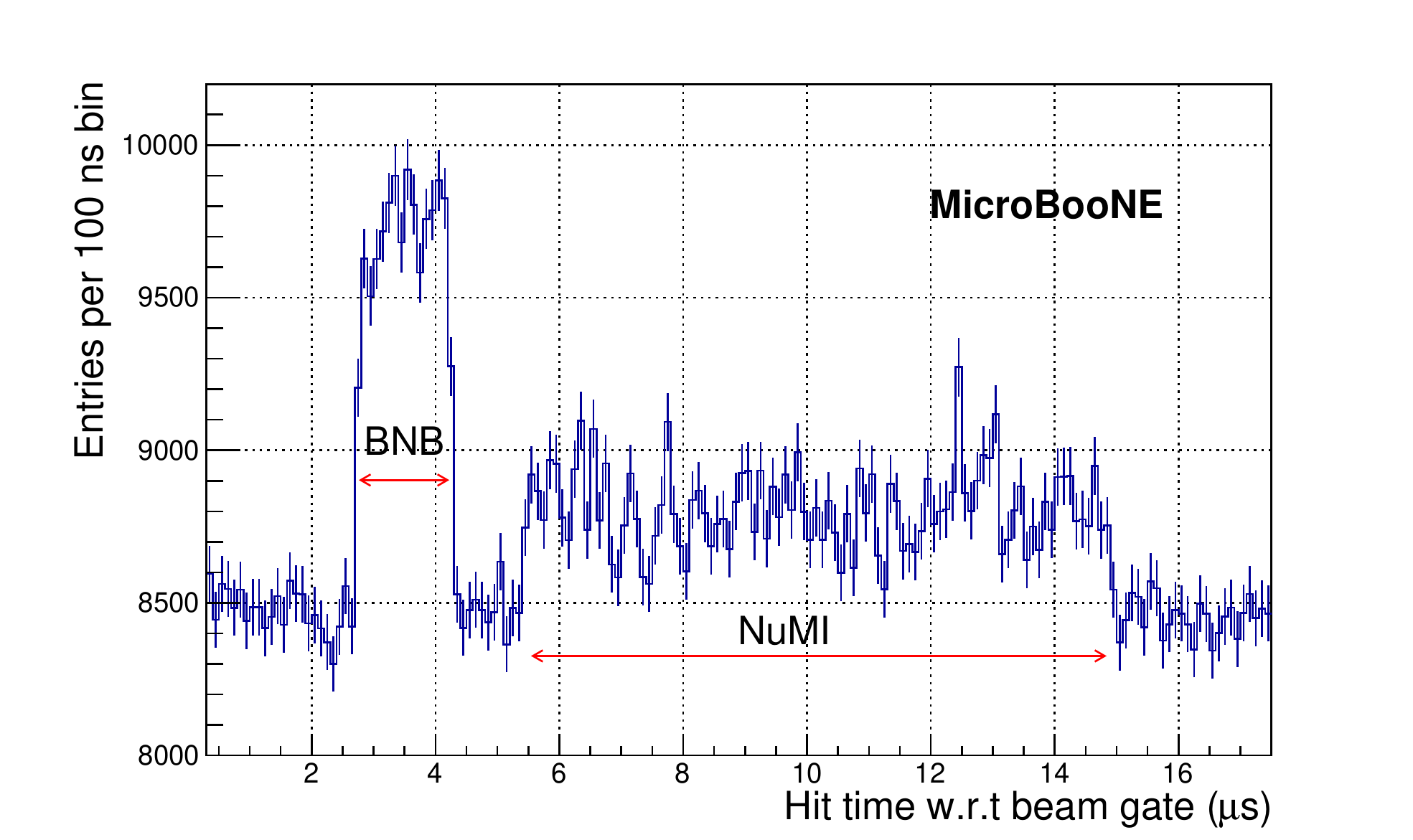}
\caption[]{MicroBooNE neutrino events seen by the CRT. Outgoing muons from $\nu_\mu$ interactions can be detected by the CRT. One can see the hits related to BNB are accumulated starting at $\sim$2.5~$\mu$s with a width of $\sim$1.6~$\mu$s. The wider distribution starting at $\sim$5~$\mu$s which extends over $\sim$10~$\mu$s is related to $\nu_\mu$ interactions in the NuMI beam.} 
\label{fig:crtevents}
\end{figure}

Outgoing muons from neutrino interactions in the LArTPC can only be identified using CRT data. Muons from $\nu_\mu$ interactions can travel outside the cryostat and intersect the CRT, producing detectable CRT hits. The CRT receives triggers from BNB and NuMI neutrino beams. These triggers are different in propagation delays but are not separated before being sent to the CRT. Two groups of beam-related overlapping CRT hits are shown in figure~\ref{fig:crtevents}. A prominent excess of CRT triggers is found starting at $\sim$3~$\mu$s with respect to the LArTPC event window and spread over $\sim$1.6~$\mu$s, which is consistent with the BNB beam spill structure. The sharp rise of the leading edge of the BNB signals indicates an absolute timing resolution of better than 100~ns for the CRT system. The wider plateau of CRT hits arising at $\sim$5~$\mu$s and spreading over a $\sim$10~$\mu$s time interval is formed by the $\nu_\mu$ neutrinos from the NuMI beam.

To remove background cosmic muons, LArTPC triggers and CRT triggers corresponding to the same beam events need to be synchronized. This matching is done by merging LArTPC and CRT events with the same GPS timestamps. 
Algorithms for matching MicroBooNE LArTPC features and CRT trajectories are under development. 

\subsection{Cosmic events}

Scintillation light induced by cosmic muons passing through CRT planes can be reconstructed as CRT hits on specific CRT channels. These hits are crucial for the higher-level reconstruction of cosmic muon trajectories. 

\begin{figure}[htb!pb]
\centering
\includegraphics[width =0.9\textwidth]{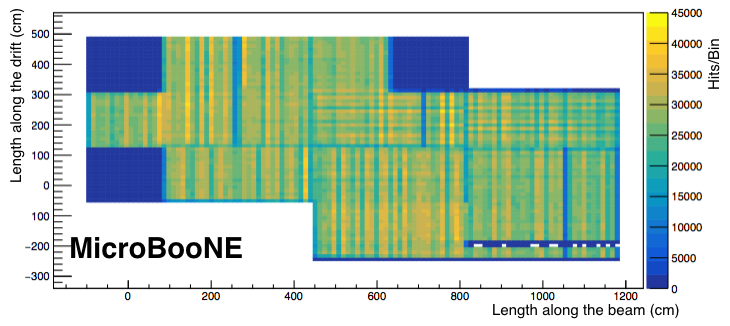}
\caption{Map of CRT hits in the top plane~(color version online). The colder strips are non-operative channels. }
\label{fig:eventstop}
\end{figure}

\begin{figure}[htb!pb]
\centering
\includegraphics[width = 0.9\textwidth]{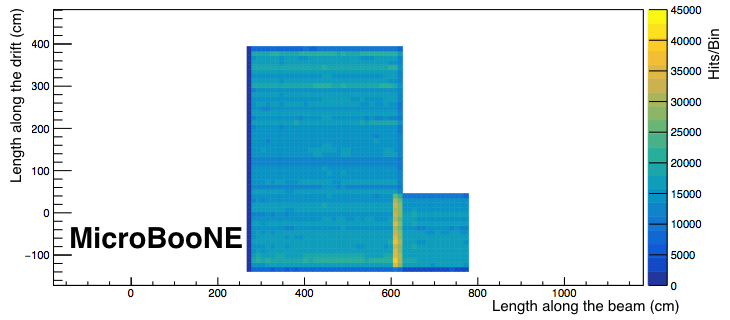}
\caption{Map of CRT hits in the bottom plane~(color version online). The overlap of modules (hotter zones) and the cryostat shadow are visible compared to figure~\ref{fig:topbottompalnes}. 
}
\label{fig:eventsbottom}
\end{figure}

\begin{figure}[htb!pb]
\centering
\includegraphics[width =0.9\textwidth]{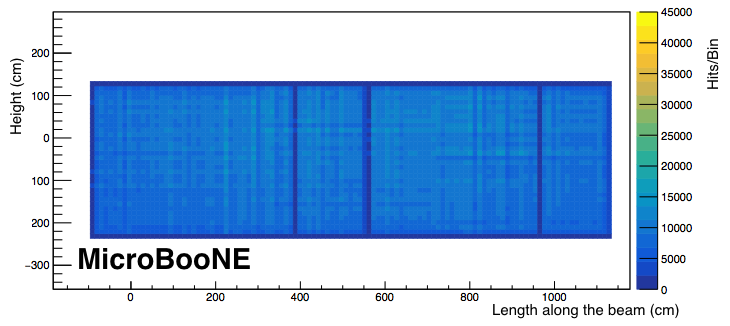}
\caption{Map of CRT hits in the feedthrough side plane~(color version online). The darker strips are non-operative channels. }
\label{fig:eventsft}
\end{figure}

\begin{figure}[htb!pb]
\centering
\includegraphics[width =0.9\textwidth]{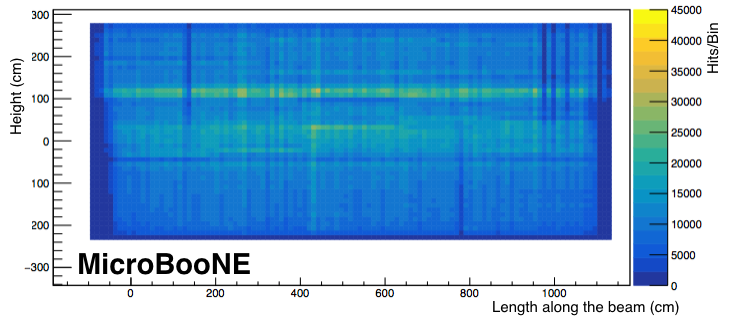}
\caption{Map of CRT hits in the pipe side plane~(color version online). The overlap of modules in the pipe side plane is visible comparing to figure~\ref{fig:sideplanes}. The long strip region in the upper plane corresponds to the area of a three-module overlay. The darker strips are non-operative channels. }
\label{fig:eventspipe}
\end{figure}

Data used here for demonstration of functionality of the CRT was collected over the course of one day of data-taking. CRT hits of cosmic muons traversing the top plane are shown in figure~\ref{fig:eventstop}. CRT hits reconstructed in the top plane are uniformly distributed. The hit map of the bottom plane is shown in figure~\ref{fig:eventsbottom}. In the bottom plane, CRT hit rates are lower in the center along the drift direction, an effect caused by the absorption of cosmic muons in the cylindrical LAr region inside the cryostat. 
CRT hit maps of the side planes are shown in figure~\ref{fig:eventsft} and figure~\ref{fig:eventspipe}. The data shows that the position information provided by CRT qualitatively reflects the muon fluxes in various locations at LArTF.


\newpage

\subsection{Timing resolution}

\begin{figure}[htb!pb]
\centering
\subfloat[$\Delta$t for CRT hits in the bottom plane.]{\label{fig:timeresolution-a}\includegraphics[width=0.495\textwidth]{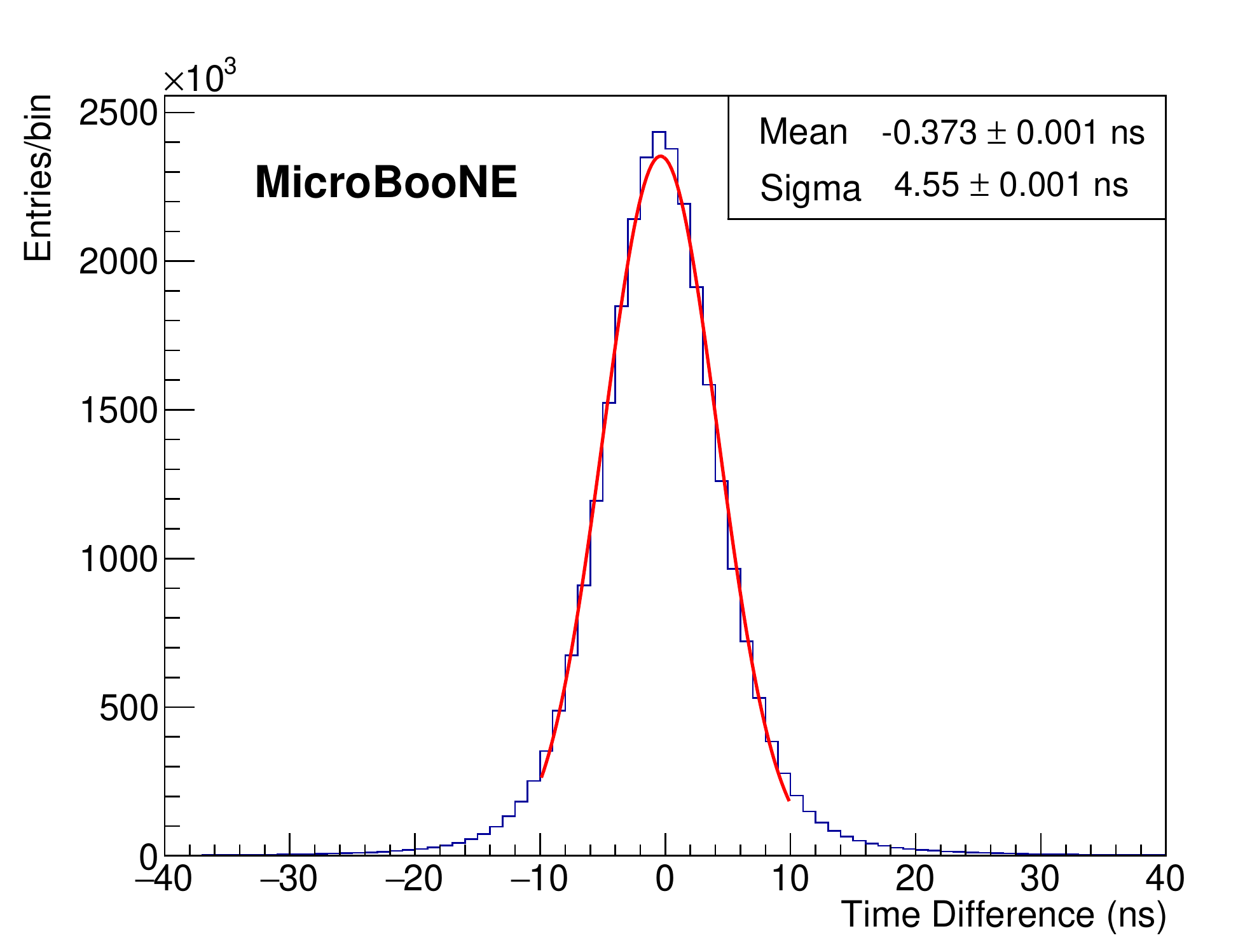}}
\hspace{\fill}
\subfloat[Zoomed-in $\Delta$t for CRT hits in the bottom plane.]{\label{fig:timeresolution-b}\includegraphics[width=0.495\textwidth]{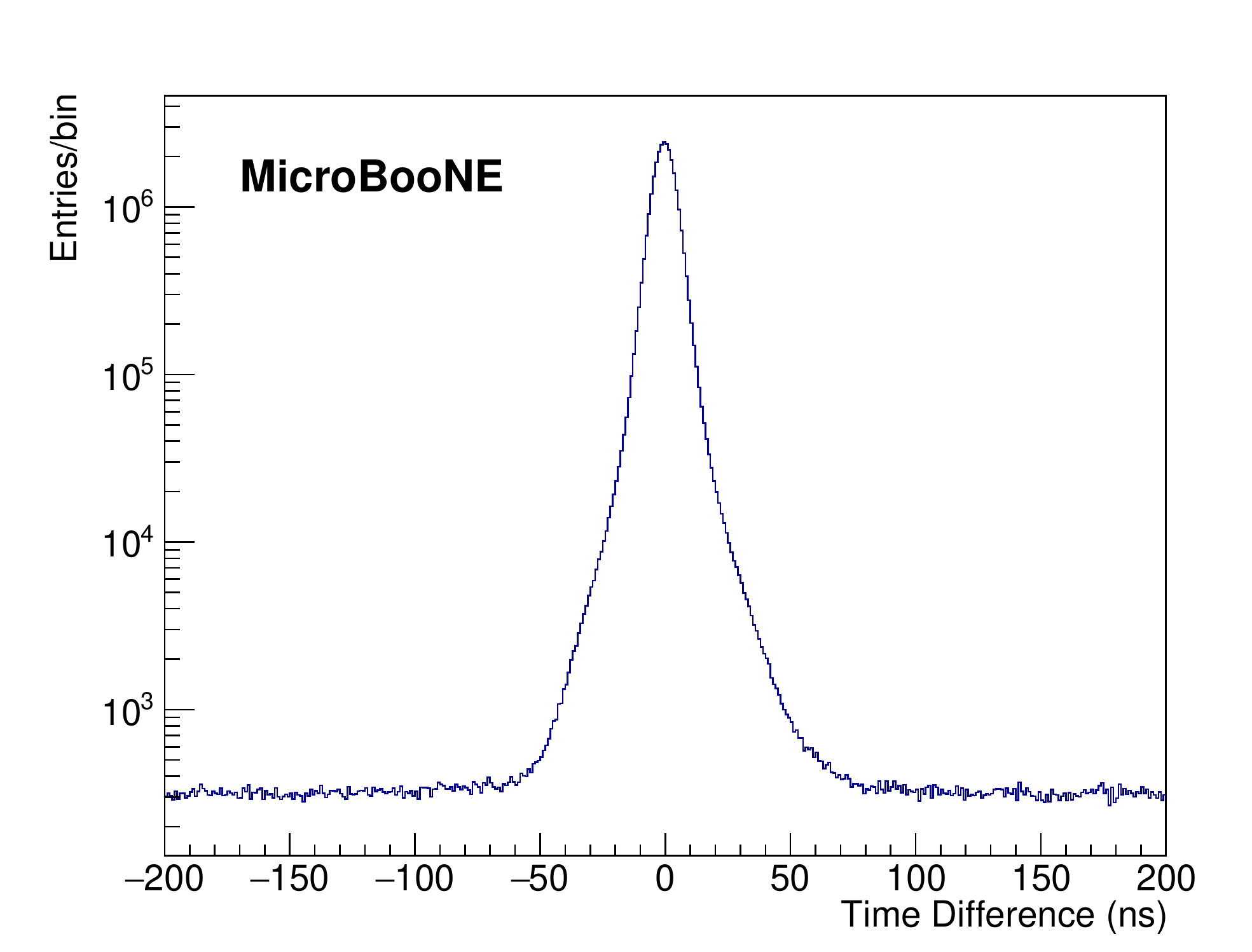}}
\caption[]{(a)~$\Delta$t between CRT hits in the two layers in the bottom plane. The width of the Gaussian fit is $\sigma=4.55$~ns, implying an average timing resolution of a single module of 4.55/ $\sqrt[]{2}=3.22$ ns, or 4.55/2=2.28~ns for a pair of XY-modules. (b)~The same distribution with a log scale and expanded in time. The random coincidence rate is very low.}
\label{fig:timeresolution}
\end{figure}

Good timing resolution is critical for matching LArTPC and CRT events. In addition to the neutrino-induced CRT hit time shown in figure~\ref{fig:crtevents}, the timing resolution of the CRT modules has been measured using the bottom plane modules. As shown in figure~\subref*{fig:timeresolution-a}, the time difference, $\Delta$t, between CRT hits of two x-y adjacent modules in the bottom plane shows a resolution of $\sim$4~ns which matches the resolution of the MicroBooNE light collection system~\cite{ub_detector}. Figure~\subref*{fig:timeresolution-b} shows that the coincident CRT hits are induced by muons with a very low rate of background coincidences.


\newpage

\section{Conclusions}
In this paper we have presented the details of the design, testing, construction, and operational status and performance of the MicroBooNE CRT external sub-system. The MicroBooNE CRT has been constructed and commissioned in two phases with phase I in July-September 2016 for the bottom, feedthrough side and pipe side planes and phase II in February 2017 for the top plane. The supporting electronic racks began operating in October 2016. MicroBooNE began CRT data-taking in March 2017. Using CRT data when MicroBooNE is recording beam data, we have demonstrated that CRT modules are capable of achieving a timing resolution on the order of $\sim$4~ns. 
We have also demonstrated the ability to reconstruct general physics features of the LArTF cosmic muon flux from BNB and NuMI neutrino beam events using only CRT data. The latter results indicate the proper functionality of the full CRT readout chain, from scintillator strip and SiPM readout, through FEB data acquisition and triggering, to proper synchronization with other LArTF and MicroBooNE timing signals. With the future development of CRT-LArTPC-PMT feature matching algorithms, the data from this fully-functional CRT system will enable new techniques mitigating backgrounds from cosmogenic particles and improve the reconstruction of LArTPC physics.

\acknowledgments

This document was prepared by the MicroBooNE collaboration using the resources of the Fermi National Accelerator Laboratory (Fermilab), a U.S. Department of Energy, Office of Science, HEP User Facility. Fermilab is managed by Fermi Research Alliance, LLC (FRA), acting under Contract No. DE-AC02-07CH11359.  MicroBooNE is supported by the following: the U.S. Department of Energy, Office of Science, Offices of High Energy Physics and Nuclear Physics; the U.S. National Science Foundation; the Swiss National Science Foundation; the Science and Technology Facilities Council of the United Kingdom; and The Royal Society (United Kingdom).  Additional support for the laser calibration system and cosmic ray tagger was provided by the Albert Einstein Center for Fundamental Physics, Bern, Switzerland.


\appendix
\section{Acronyms}

\begin{table}[htb!pb]
\begin{center}
\caption{List of acronyms.}
\begin{adjustbox}{}
\begin{tabular}{| l | l | c | c | c | c | c | c | c |}
\hline
Acronym& Explanation\\ 
\hline
ADC & Analog to Digital Converter \\
A.U. & Arbitrary Unit \\
BNB & Booster Neutrino Beam\\
CRT & Cosmic Ray Tagger\\
DAB & DZero Assembly Building \\
DAQ & Data Acquisition \\
EPO &  Emergency Power Off \\
EVB & Event Building \\
FEB & Front End Board \\
FPGA & Field-Programmable Gate Array \\
LArTF & Liquid Argon Test Facility \\
LArTPC & Liquid Argon time Projection Chamber \\
LHEP & Laboratory for High Energy Physics \\
NuMI & Neutrino at the Main Injector \\
PDU & Power Distribution Unit \\
PMT & Photomultiplier Tubes \\
RPS & Rack Protection System \\
SBN & Short Baseline Neutrino (program)\\
SiPM &  Silicon Photomultipliers \\
TIN & Trigger In\\
TOUT & Trigger Out \\
WLS & Wavelength Shifting \\
UPS & Uninterrupted Power Supply \\

\hline
\end{tabular}
\end{adjustbox}
\label{T:acronym}
\end{center}
\end{table}

\newpage




\end{document}